\documentclass[review]{elsarticle}
\usepackage{lineno,hyperref}
\usepackage{xcolor}
\hypersetup{colorlinks=true,linkcolor=red,citecolor=green,urlcolor=blue}
\usepackage{caption}

\usepackage{geometry}
\usepackage{amsmath}
\usepackage{tabularx,array}
\usepackage{multirow}
\usepackage{subcaption}

\begin{document}


\begin{frontmatter}
\title{Cooperation Dynamics with Environmental Feedback: Steady States and Transitions}

\author{Yu Chen}

\author{Ji-Qiang Zhang}

\author{Bin-Quan Li\corref{mycorrespondingauthor}}
\cortext[mycorrespondingauthor]{Corresponding author}
\ead{lbqdzh@163.com}

\address{School of Physics, Ningxia University, Yinchuan, Ningxia 750021, China}

\begin{abstract}The evolutionary steady states and regulatory mechanisms of cooperative behavior constitute core research themes in complex systems and evolutionary game theory. Within natural populations, bidirectional feedback emerges between individual strategy updates and endogenous environmental payoff structures, which fundamentally shapes the evolutionary trajectories of group cooperation. Building upon the classical evolutionary game framework, this paper develops a strategy–payoff environmental-feedback dynamical model encompassing four canonical game paradigms: the Prisoner's Dilemma, the Snowdrift Game, the Stag-Hunt Game, and the Harmony Game. Via rigorous mathematical derivations, we obtain a standard canonical replicator equation that is fully analytically equivalent to the original high-dimensional environment-coupled system. Integrating bifurcation diagrams, phase-plane portraits, and numerical verification, this study systematically investigates the joint regulatory impacts of initial cooperation fraction, temptation to defect, and punishment payoff on cooperative evolutionary dynamics. The results reveal that the Stag-Hunt Game and Harmony Game both display prominent bistability, with long-run evolutionary outcomes strongly dependent on initial states. By contrast, the Prisoner’s Dilemma exhibits robust evolutionary behavior, where mixed-strategy equilibria remain permanently stable. The interaction between temptation and punishment drives sequential cross-type game transitions at multiple critical dynamical thresholds. Notably, within this environment-feedback coupled system, sufficient punishment does not merely curb defection incentives; it shifts critical dynamical bifurcation thresholds and delays the destabilization of cooperative steady states, further reshaping these sequential cross-type game transitions. Numerical examinations verify complete self-consistency and identical dynamical characteristics between the original environment-coupled system and its analytically equivalent canonical form, further clarifying the stability switching rules for all equilibria. Furthermore, outputs from spatial lattice simulations are qualitatively consistent with theoretical predictions derived from the canonical replicator framework, which corroborates the robustness and generalizability of the proposed environmental-feedback theoretical framework. This work identifies the intrinsic regulatory mechanisms governing cooperative evolution arising from endogenous payoff feedback and initial-condition sensitivity, providing solid theoretical foundations for explaining and optimizing collective prosocial behaviors within complex social systems.
\end{abstract}

\begin{keyword}
Feedback-evolving game \sep Cooperative evolution \sep Social dilemma \sep Replicator equation \sep Environmental feedback
\end{keyword}

\end{frontmatter}


\section{Introduction}
Cooperation is a prosocial behavior that benefits others at the expense of individual interests, reducing the donor's fitness while enhancing the recipient's survival and competitive advantages \cite{han2026cooperation,hardin1968tragedy,nowak2006five}.
It is widespread in natural multi-level systems and human societies; for example, under starvation conditions, social amoebae ensures population survival and reproduction through aggregation, cell differentiation, altruistic stalk cell apoptosis, and coordinated regulation \cite{weitz2016oscillating,rand2011dynamic}. In social insect colonies, worker ants and bees voluntarily forfeit reproductive rights to serve the queen, relying on high-level collaboration to secure population continuity. However, altruistic cooperation fundamentally conflicts with Darwinian evolution: under natural selection, individuals generally seek to maximize their own benefits and personal fitness, which raises the key scientific question of how cooperation can spontaneously emerge and remain stable in a population of rational, self-interested individuals.

To address this question, evolutionary game theory provides a crucial framework for studying group strategy evolution and cooperation dynamics. By elucidating strategic interactions and the survival of the fittest, researchers have systematically investigated classic models such as the Prisoner's Dilemma \cite{tilman2020evolutionary,ito2024complete}, the public goods game \cite{chen2018punishment,li2025public}, and the Snowdrift Game \cite{doebeli2005models,hauert2004spatial}.
In the traditional two-strategy framework, players choose either cooperation or defection; the payoff structure of each pairwise combination defines different game types, each with distinct evolutionary trends and steady-state characteristics. Replicator dynamics serves as the classical deterministic modeling approach for strategy evolution in well-mixed populations \cite{wang2020eco,szolnoki2017environmental}. Using the distinct payoff structures of the Prisoner's Dilemma \cite{bravetti2018optimal}, Snowdrift Game \cite{doebeli2005models,chen2017evolutionary}, Harmony Game \cite{kiraly2026evolutionary,yang2020ranking}, and Stag-Hunt Game \cite{roca2009evolutionary,pacheco2008evolutionary}, replicator dynamics quantitatively reveals the evolutionary dynamics of strategies—ranging from defection dominance and minority advantage to bistability and natural cooperative advantage—forming the core theoretical foundation for explaining the emergence, maintenance, and collapse of cooperation. Furthermore, multi-game environments have emerged as a critical research frontier in evolutionary game theory. Early studies predominantly focused on static one-shot game scenarios, while the academic community has increasingly turned its attention to hybrid game frameworks, which can better characterize the dynamic heterogeneity of real-world interactions. Previous work has investigated strategic dominance in two types of hybrid games, including games with fixed and stochastically varying game compositions. Through rigorous mathematical analysis and computational simulations, we demonstrate that the criterion for strategic dominance under weak selection is equivalent to that for average individual one-shot games. This equivalence establishes a unified framework for simplifying complex multi-game interactions and advances the mechanistic understanding of cooperative evolution in structured populations \cite{lv2024fixation}.

Based on this foundation, the academic community has proposed several cooperation‑-promoting mechanisms within the framework of evolutionary game theory, primarily including reward and punishment systems \cite{wang2025coevolutionary,li2021egoistic,rockenbach2006efficient,mcavoy2019asymmetric}, individual migration and diffusion \cite{takano2015environmentally,traulsen2006evolution,santos2008social}, reputation evaluation frameworks \cite{mcnamara2015reputation,wang2020social,jia2025evolutionary,yang2018promoting}, and strategic coevolution \cite{hilbe2018evolution,li2014evolution,hebertdufresne2017strategic}. Most previous studies assume a static external environment that does not evolve with population behavior. However, this idealized premise differs significantly from real-world conditions. Both natural systems and social contexts exhibit dynamic and time‑-varying characteristics: individual strategic choices continuously reshape public environments and resource conditions, while evolving environmental states in turn constrain individual payoffs and decision‑-making processes \cite{ma2023evolution}. This interaction forms a bidirectional feedback relationship directly reflected in the dynamic adjustment of the game payoff matrix \cite{yang2020environmental,ma2023evolution}. Consequently, the environmental feedback mechanism has become a crucial framework for analyzing the evolution of cooperation, attracting extensive attention from researchers worldwide in recent years\cite{gao2024evolutionary,weitz2016oscillating,szolnoki2017environmental,chen2018punishment,hauert2019asymmetric,tilman2020evolutionary,stella2023irrational,stella2022lower,lu2023antisocial,hua2023facilitating,liu2024fostering,lu2024reciprocal,li2022alternating}.
.

Within this rapidly developing field, a generalized replicator dynamics model integrating environmental feedback was developed \cite{weitz2016oscillating}, emphasizing defection constraints and cooperative incentives to mitigate the tragedy of the commons. A strategy-payoff coevolutionary model for the Prisoner's Dilemma was constructed \cite{szolnoki2017environmental}, showing that co-feedback raises cooperation rates regardless of network topology and slows the transition to cooperation. Governance strategies for public resources using punishment were investigated \cite{chen2018punishment}, and an asymmetric evolutionary game framework for heterogeneous individuals was proposed \cite{hauert2019asymmetric}. A comprehensive model of environmental feedback games was systematically established \cite{tilman2020evolutionary}, laying a theoretical foundation. Later work incorporated irrational behavior, network topology, antisocial behavior, and adaptive punishment. The effects of irrational decisions and node degree on cooperation stability were examined \cite{stella2023irrational,stella2022lower}, and environmental feedback rules linked to antisocial behavior were introduced \cite{lu2023antisocial}, revealing how negative behaviors interfere with the spatial Prisoner's Dilemma. An adaptive feedback punishment strategy was adopted \cite{hua2023facilitating}, confirming that coupling altruistic punishment with the environment promotes cooperation, especially in public goods games. Dynamic environmental feedback in public goods games was introduced \cite{liu2024fostering}, enabling punishment intensity to adjust with the proportion of cooperators and balancing individual and collective interests for long-term stability. Environmental feedback was incorporated into spatially structured populations \cite{gao2024evolutionary}, demonstrating a critical temptation threshold above which cooperation dominates. Finally, a reciprocal reward-based environmental feedback mechanism was proposed \cite{lu2024reciprocal}, providing a new approach to sustain cooperation in spatial games.

In addition, some studies have constructed time-varying evolutionary models in which the payoff matrix adaptively adjusts according to cooperation levels, incorporating finite-population stochastic dynamics \cite{yang2018promoting,wang2019spontaneous} and environment-strategy bidirectional coupling rules \cite{quan2020social}. Lv et al. \cite{lv2024fixation} applied the Moran stochastic process to introduce environmental feedback into finite-population games, establishing a time-varying payoff evolutionary model. They systematically explored the effects of population size, selection intensity, and feedback time scale on the fixation probability of cooperation. This line of research has improved stochastic evolution theory for finite populations and broadened the application scenarios of environmental feedback. Nevertheless, critical limitations remain: heavy reliance on the stochastic framework restricts global parameter partitioning and dynamic mechanism analysis in deterministic replicator dynamics systems. Furthermore, the focus on statistical features of cooperation evolution ignores systematic quantitative investigations of fixed-point distribution, stability switching, and critical dynamic evolution conditions within the system.

In contrast to most prior literature that relies on stochastic evolutionary frameworks, the present work adopts classical deterministic replicator dynamics as the core analytical tool. We concentrate on how key payoff parameters modulate cooperative steady states and systematically explore the corresponding evolutionary rules and dynamic responses as parameters vary. Special attention is paid to the global structure of the parameter space, single-parameter evolutionary trajectories, and the internal mechanisms governing local equilibrium stability. This research fills a notable research gap: comprehensive global dynamic analysis for environmental feedback games under a deterministic setup remains underexplored in existing studies.

The remainder of this paper is structured as follows. Section~\ref{section.2} constructs the replicator dynamic framework, derives governing differential equations, and solves all system fixed points together with their feasible existence constraints. Section~\ref{section.3} conducts thorough theoretical stability analysis, partitions the full parameter domain into distinct dynamic regimes, establishes universal stability criteria, and uncovers critical bifurcation points and equilibrium switching mechanisms. Section~\ref{section.4} cross-validates all analytical predictions via multi-faceted numerical evidence, including global parameter heatmaps, local phase-plane portraits, and time-series trajectories of the cooperation fraction, covering macroscopic global patterns, continuous transition processes, and microscale stability mechanisms. Finally, Section~\ref{section.5} concludes the whole work by summarizing core evolutionary regularities of cooperation and puts forward prospective extensions including model improvement, real-scenario generalization, and multi-dimensional dynamic expansion.

\section{Model}
\label{section.2}
In this section, we construct an evolutionary game framework embedded with real-time environmental feedback. Most existing studies adopt lagged coupling relations between historical environmental states and population strategy distributions. In contrast, our model assumes that environmental quality is solely determined by the instantaneous proportion of cooperators within the population, independent of past behavioral records.

Consider a finite population of total size $N$, where $i$ denotes the number of cooperative individuals. The cooperation frequency is defined as $x = i/N$. Under this setting, environmental conditions and individual strategic choices evolve on identical timescales, and the entire system satisfies the memoryless Markov property for both strategy updating and environmental evolution. All individuals interact via symmetric two-player games with two pure strategies: cooperation ($C$) and defection ($D$). The time-varying payoff matrix $U(i)$ is constructed as a convex linear combination of two payoff structures corresponding to two extreme monomorphic population states: the optimal fully cooperative environment ($i=N$) and the degraded full-defection environment ($i=0$). This construction accurately captures continuous smooth variations in environmental quality as the cooperation frequency changes.

\subsection{Payoff Structure of the Optimal Cooperative Environment}
When the population reaches full cooperation $i=N$ and the environment remains intact, we adopt the standard four-tuple payoff framework for social dilemmas. If two individuals both choose cooperation, each obtains the mutual cooperation reward $R$. When both players defect simultaneously, each receives the mutual defection punishment payoff $P$. For mixed pairs where one individual cooperates and the other defects, the cooperator gets the sucker payoff $S$, whereas the defector gains the temptation payoff $T$.

Under such intact environmental conditions, short-term individual self-interest drives the overexploitation of public resources, and unilateral defection becomes the dominant strategy, rendering mutual defection the unique Nash equilibrium. Persistent widespread defection continuously depletes public resources and gradually pushes the entire system toward environmental degradation.

\subsection{Reversed Payoff Structure under Degraded Environmental Conditions}
When the entire population consists of defectors ($i=0$) and the environment undergoes severe degradation, the extra short-term gains generated by defection vanish entirely. To capture the real-world logic that only collective cooperation can restore depleted resources, we fully reverse the payoff mapping for all game interactions relative to the benchmark payoff matrix under intact environmental conditions.

Simple uniform rescaling or additive shifts of payoff entries cannot alter their ordinal ranking, so such manipulations are incapable of switching the dominant strategy from defection to cooperation. Payoff permutation is therefore required to realize the qualitative inversion of individual-level strategic incentives. This full-reversal permutation is chosen for parametric parsimony: it reuses the primitive payoff quadruple $\{T,R,P,S\}$ from the intact-environment benchmark and avoids introducing additional free parameters that would obscure comparative analysis across environmental regimes. We note that this construction represents an idealized benchmark; real-world socio-ecological systems may exhibit only partial incentive inversion upon environmental degradation.

When both players cooperate in a degraded environment, each participant receives the temptation payoff $T$ defined for the intact environment. If both individuals choose defection, their payoff reduces to the original sucker's payoff $S$. For mixed interaction pairs, cooperators matched with defectors obtain only the original mutual-defection punishment payoff $P$, whereas defectors exploiting cooperative partners earn the original mutual-cooperation reward $R$.

Under this reversed-payoff framework, over-exploitation no longer yields extra individual benefits, and mutual cooperation becomes the unique Nash equilibrium capable of restoring environmental quality. These two sets of benchmark payoff rules jointly constitute a dual-state switching mechanism, laying the foundation for bidirectional feedback between individual strategies and environmental evolution \cite{lv2024fixation}.

The population's fraction of cooperators dynamically modulates the global payoff matrix and further shapes the evolutionary trajectories of environmental states. Meanwhile, adaptive shifts in environmental payoff structures alter individual strategic payoffs and constrain the competitive evolution between cooperation and defection. This reciprocal interaction establishes an endogenous synergistic coupling between strategy evolution and environmental restoration.
\begin{equation}
\begin{aligned}
U(i) &= \left(1-\frac{i}{N}\right)
\begin{pmatrix}
T & P \\
R & S
\end{pmatrix}
+ \frac{i}{N}
\begin{pmatrix}
R & S \\
T & P
\end{pmatrix} \\
&=
\begin{pmatrix}
\displaystyle \frac{i}{N}(R-T)+T & \displaystyle \frac{i}{N}(S-P)+P \\[6pt]
\displaystyle \frac{i}{N}(T-R)+R & \displaystyle \frac{i}{N}(P-S)+S
\end{pmatrix}
\end{aligned}
\label{eq.1}
\end{equation}

The primitive payoff quadruple uniquely defines two time-invariant payoff matrices corresponding to the extreme monomorphic population states $x=0$ and $x=1$. Accordingly, the Stag-Hunt (SH), Prisoner's Dilemma (PD), Harmony Game (HG), and Snowdrift Game (SD) regions delineated on the $T-P$ parameter plane in Figure~\ref{fig.1} only describe the benchmark game classes linked to these two limiting monomorphic configurations, rather than fixed social dilemma incentives that hold for all interior cooperation fractions $x\in(0,1)$. The composite payoff matrix adopted in this work is expressed as a convex linear combination of the two extreme monomorphic payoff matrices. As the cooperation fraction $x$ evolves, the ordinal ranking of pairwise payoff entries changes endogenously, meaning the instantaneous strategic incentive structure faced by interacting agents cannot be uniquely determined merely by the static ordering of the primitive payoffs. To resolve this apparent conceptual contradiction, we propose a dual-layer classification paradigm. Static partitioning on the $T-P$ plane acts as a macroscopic parameter screening tool that captures the asymptotic evolutionary tendencies induced by the two boundary monomorphic payoff matrices. Meanwhile, the closed-form transition threshold $x_{\text{trans}}=(P-S)/[(P-S)+(R-T)]$ quantifies the critical cooperation level at which payoff orderings reverse. A cross-type game transition occurs along the evolutionary trajectory if $0<x_{\text{trans}}<1$; in contrast, the static benchmark classification remains uniform over the entire state space whenever $x_{\text{trans}}\notin(0,1)$. Static zoning and dynamic threshold analysis serve distinct analytical purposes and are mutually consistent. All figure captions are revised to clearly specify that the plotted domains correspond exclusively to limiting monomorphic benchmark games, rather than time-invariant game classes valid over the full interval $x\in(0,1)$.

In a population of size $N$, the expected returns of the cooperator and the defector with $i$ cooperators are respectively:
\begin{equation}
\begin{aligned}
\varPi_C(x) &= x U_{CC} + (1-x) U_{CD} \\
&= x\big[T + x(R-T)\big] + (1-x)\big[P + x(S-P)\big]
\label{eq.2}
\end{aligned}
\end{equation}

\begin{equation}
\begin{aligned}
\varPi_D(x) &= x U_{DC} + (1-x) U_{DD} \\
&= x\big[R + x(T-R)\big] + (1-x)\big[S + x(P-S)\big]
\label{eq.3}
\end{aligned}
\end{equation}

With the linear convex dynamic payoff matrix and Markov coevolution rule fully defined, we further derive the one-dimensional simplified replicator dynamic equation in the subsequent section. On this basis, we solve all equilibrium fixed points and evaluate their local asymptotic stability through Jacobian linearization, as well as analyze the transcritical bifurcation behaviors induced by varying the defection temptation parameter $T$.

\section{Theoretical results}
\label{section.3}
This section presents a systematic theoretical analysis of steady‑state existence, local stability, and global dynamical evolutionary structure for the proposed environmental feedback evolutionary game model. By solving equilibrium conditions and evaluating the local derivative of the dynamical function at each fixed point, we derive the existence ranges, stability properties, and evolutionary attraction features of each fixed point. In addition, we clarify multiple dynamical evolutionary scenarios modulated by core payoff parameters. Using the constraint relations among key dimensionless payoff parameters, we rigorously classify the four canonical game types and quantitatively identify the critical parameter thresholds corresponding to distinct dynamical evolutionary regimes. Furthermore, we uncover the inherent rules governing stability switching mechanisms and continuous game-type transitions, and characterize the global dynamical landscape of cooperative evolution under continuous parameter variation.

\subsection{Replication dynamics equation}
Under the infinite population approximation, the continuous-time evolution of the strategy frequency is described by replication dynamics, and its form is:
\begin{equation}
\dot{x} = x(1-x)\left(\varPi_C - \varPi_D\right)
= x(1-x)(2x-1)\left[(S-P) + x(R-T-S+P)\right]
\label{eq.4}
\end{equation}

\subsection{Solve for the fixed point}
Setting the time derivative of evolutionary dynamic Equation~(\ref{eq.4}) to zero yields the system's equilibrium condition, from which we further derive all steady-state fixed points of the model. The system admits three classes of steady-state solutions. The first class corresponds to boundary fixed points, representing extreme evolutionary outcomes where the population is dominated by a single strategy. The second class comprises a parameter-independent interior fixed point whose equilibrium value exhibits strong robustness against core payoff control parameters. The third class consists of parameter-dependent interior fixed points, whose equilibrium positions vary continuously with the key payoff parameters. The existence and local stability of these fixed points are strictly constrained by feasible parameter intervals. Ultimately, we obtain four equilibrium fixed points, among which two are interior fixed points:
\begin{equation}
\begin{aligned}
&x_1^* = 0,\\
&x_2^* = 1,\\
&x_3^* = \frac{1}{2},\\
&x_4^* = \frac{P-S}{P-S+R-T}
\end{aligned}
\label{eq.5}
\end{equation}

\subsection{Verify whether the fixed point exists}
Fixed points stand for equilibrium states of the evolutionary dynamical system and govern the long-run convergence trend of population strategy frequencies. This section first solves all boundary and interior mixed fixed points of the system using the evolutionary equilibrium condition $\mathrm{d}x/\mathrm{d}t = 0$. Boundary fixed points correspond to pure-strategy equilibria: specifically, $x_1^* = 0$ stands for the all-defection steady state, whereas $x_2^* = 1$ represents the full-cooperation steady state. Interior fixed points describe mixed-strategy equilibria where cooperation and defection coexist, and their feasibility is restricted by the payoff parameters $R, S, T, P$. Their biological interpretability and validity are limited to specific parameter intervals. By combining parameter constraints and domain bounds, we can explicitly formulate the existence criteria for all fixed points. Pure-strategy boundary fixed points exist over the full parameter space and constitute the fundamental equilibrium states for all game classes. In contrast, interior mixed fixed points are subject to stringent parameter restrictions. They fall within the valid strategy interval $(0,1)$ only if payoff differences satisfy certain inequalities. For some parameter regimes, only one interior fixed point exists, or both interior fixed points vanish at once, leaving the system with solely pure-strategy steady states and simplifying its overall dynamical structure. The local stability of each fixed point is determined by the linearized derivative of replicator dynamics evaluated at the equilibrium, defined as
\begin{equation}
J = \left.\frac{\mathrm{d}}{\mathrm{d}x}\left(\frac{\mathrm{d}x}{\mathrm{d}t}\right)\right|_{x=x^*}
\label{eq.6}
\end{equation}
where $J<0$ indicates local asymptotic stability, whereas $J>0$ indicates instability. The boundary fixed points $x_1^*=0$, $x_2^*=1$ and the interior fixed point $x_3^*=1/2$ exist globally, while the parameter‑dependent interior fixed point $x_4^*$ exists if and only if $0<x_4^*<1$. The corresponding stability conditions are re‑derived and summarized as follows: $x_1^*=0$ is stable when $P<S$; $x_2^*=1$ is stable when $R>T$; the interior fixed point $x_3^*=1/2$ is stable provided $R+S-P-T<0$; and the interior fixed point $x_4^*$ is stable when $0<x_4^*<1$ and $f'(x_4^*)<0$. Characterizing the existence and stability of all fixed points allows us to preliminarily classify the fundamental structural properties of the system, which lays a theoretical foundation for subsequent local stability analysis, game‑type classification, and dynamical evolutionary behavior analysis.

\subsection{Classification of classic game types}
(1) The Harmony Game

When parameters satisfy $S > P$ and $R > T$, the system falls into the Harmony Game regime \cite{kiraly2026evolutionary,yang2020ranking}. This regime contains no severe social dilemma, allowing cooperative and defective strategies to persist independently. Stability analysis reveals that both boundary fixed points—the full-cooperation state $x_2^*=1$ and the all-defection state $x_1^*=0$—are locally asymptotically stable. The system displays characteristic bistability. The parameter-dependent interior fixed point $x_4^*$ lies outside the feasible strategy interval $[0,1]$ under Harmony-Game parameter conditions and therefore cannot serve as a separatrix. Instead, the interior fixed point $x_3^*=1/2$ is unstable and acts as the separatrix dividing the basins of attraction of the two boundary equilibria. Evolution within this parameter regime exhibits prominent path dependence. If the initial cooperation fraction is sufficiently large (above $x_3^*=1/2$), trajectories converge to the full-cooperation steady state. If defectors dominate the initial population (below $x_3^*=1/2$), the system evolves toward the all-defection equilibrium. The two pure-strategy steady states coexist as distinct local attractors, with no tendency for the system to converge to a mixed-strategy equilibrium from generic initial conditions.

(2) The Snowdrift Game

When parameters satisfy $S > P$ and $T > R$, the system corresponds to the Snowdrift Game \cite{doebeli2005models,chen2017evolutionary}. Here, core payoff parameters govern the system's equilibrium behavior, and the stability of steady-state fixed points is tightly constrained by parameter configurations. The boundary fixed point $x_1^* = 0$ is locally asymptotically stable, while $x_2^* = 1$ is unstable. The two interior fixed points $x_3^*=1/2$ and $x_4^*$ can exchange their stability via a bifurcation: when $S-P > T-R$, $x_3^*=1/2$ is unstable and $x_4^*$ is locally asymptotically stable; when $S-P < T-R$, $x_3^*=1/2$ becomes locally asymptotically stable while $x_4^*$ turns unstable. The equilibrium value $x_3^*=1/2$ is insensitive to core payoff parameters and thus exhibits strong parametric robustness whenever it is stable. In each parameter sub-regime, the unstable interior fixed point acts as the separatrix that divides the basins of attraction between the locally stable all-defection state $x_1^*=0$ and the locally stable interior mixed-strategy equilibrium. If the initial cooperation fraction lies below this separatrix threshold, the system converges to the all-defection steady state $x_1^*=0$; otherwise, it converges toward the locally stable interior mixed-strategy equilibrium. The system therefore exhibits bistability between the all-defection state and a mixed-cooperation equilibrium within the Snowdrift-Game parameter regime. Neither the all-defection state nor the mixed-strategy equilibrium acts as a global attractor.

(3) The Prisoner's Dilemma Game

When parameters satisfy $P > S$ and $T > R$, the regime belongs to the Prisoner's Dilemma \cite{bravetti2018optimal}, a canonical social dilemma where individual rationality yields collectively irrational outcomes. Stability analysis demonstrates that both boundary fixed points $x_1^*=0$ and $x_2^*=1$ are unstable. The parameter-dependent interior fixed point $x_4^*$ falls outside the feasible interval $(0,1)$ and thus has no biologically meaningful solution within this parameter range. Only the interior fixed point $x_3^*=1/2$ is locally asymptotically stable. Under this game setup, trajectories starting from any interior initial condition within $(0,1)$ converge toward this mixed-strategy equilibrium. Pure-cooperation states cannot persist under evolutionary dynamics, and the population evolves toward the intermediate mixed-cooperation steady state rather than full defection.

(4) The Stag-Hunt Game

When parameters satisfy $P > S$ and $R > T$, the system corresponds to the Stag-Hunt Game \cite{roca2009evolutionary}. A defining property of this game is that collective payoffs from cooperation far exceed the short-term gains of defection, aligning with the evolutionary logic of ``cooperation priority". Within this parameter region, the boundary equilibrium $x_2^*=1$ remains locally asymptotically stable and $x_1^*=0$ remains unstable. The two interior fixed points $x_3^*=1/2$ and $x_4^*$ undergo stability exchange controlled by the relative magnitude of $R-T$ and $P-S$. When $R-T < P-S$, $x_3^*=1/2$ becomes locally asymptotically stable while $x_4^*$ acts as an unstable separatrix; when $R-T > P-S$, $x_3^*=1/2$ turns unstable and $x_4^*$ becomes the locally asymptotically stable mixed-strategy equilibrium. The system maintains bistability between the full-cooperation state $x_2^*=1$ and a stable mixed-cooperation equilibrium. This structural property enables the system to sustain high cooperation levels for initial conditions above the unstable threshold and effectively prevents cooperation collapse driven by individual self-interest for sufficiently large initial cooperation fractions.

\subsection{Global Dynamic Evolution and Stability Switching}
To uncover the continuous phase-transition rules driven by tunable payoff parameters, we select the temptation to defect $T$ as the primary bifurcation control variable and perform comparative analyses at two punishment levels $P=2$ and $P=0$, with $R=3$ and $S=1$ held fixed for these test cases. For given payoff parameters, $T_{c1}$ and $T_{c2}$ denote the magnitude-ordered first and second transcritical-bifurcation thresholds. The punishment payoff $P$ is confirmed as a core regulatory parameter that reshapes the global dynamical landscape and triggers endogenous transitions between distinct game types.

When $P=2$, under the fixed payoff setting $R=3$ and $S=1$, the system yields separate evolutionary regimes separated by critical thresholds $T_{c1}=2$ and $T_{c2}=3$. For $T\in(0,T_{c1})$, the unstable boundary equilibrium $x_{1}^{*}=0$, stable boundary equilibrium $x_{2}^{*}=1$, together with two interior fixed points coexist to produce bistable dynamics, and long-run evolutionary outcomes depend entirely on initial cooperation fractions. For $T\in(T_{c1},T_{c2})$, only the two interior fixed points $x_{3}^{*}=1/2$ and $x_{4}^{*}$ undergo stability exchange. Among boundary equilibria, $x_{1}^{*}=0$ remains unstable and $x_{2}^{*}=1$ remains stable throughout this interval; no stability switch occurs for boundary fixed points. When $T>T_{c2}$, the parameter-dependent interior equilibrium leaves the feasible interval and vanishes, and both boundary fixed points lose their stability.

When $P=0$, keeping $R=3$ and $S=1$ unchanged for this test case, the two critical transcritical bifurcation points lie at $T_{c1}=3$ and $T_{c2}=4$. The system falls within the Harmony Game regime for $T\in(0,T_{c1})$, in which both boundary equilibria $x_{1}^{*}=0$ and $x_{2}^{*}=1$ are locally asymptotically stable and populations maintain high cooperation levels. For $T\in(T_{c1},T_{c2})$, the system transitions to the Snowdrift Game; here $x_{1}^{*}=0$ remains locally stable while $x_{2}^{*}=1$ loses stability. The interior fixed points $x_{3}^{*}=1/2$ and $x_{4}^{*}$ coexist within $(0,1)$ and exhibit bistable dynamics. When $T>T_{c2}$, $x_{3}^{*}=1/2$ and $x_{4}^{*}$ undergo another stability exchange. The parameter-dependent interior equilibrium $x_{4}^{*}$ still stays inside the feasible interval $[0,1]$. The system remains within the Snowdrift-Game regime, bistable dynamical characteristics persist and evolutionary outcomes are sensitive to initial cooperation fractions.

Comparative analysis of these two parameter sets verifies the validity of the simplified low-dimensional model and the precision of the stability criteria, and further elucidates the inherent evolutionary logic of cooperation, which delivers theoretical guidance for designing practical governance strategies. All regime transitions originate from the bidirectional coupling between payoff structures and environmental feedback: abundant cooperation elevates environmental quality and boosts cooperative payoffs, whereas prevalent defection degrades the environment and diminishes the relative payoffs of cooperation. This reciprocal feedback loop serves as the central driving force of the entire evolutionary system.

To summarize this subsection: punishment payoff $P$ modifies the magnitude and sequential order of the two transcritical-bifurcation thresholds and reshapes global dynamical landscapes. Variation in the temptation-to-defect parameter $T$ triggers interior or boundary fixed-point collisions and associated stability exchange events, inducing sequential game-regime transitions. Bistable behaviour and initial-condition sensitivity constitute key dynamical features for multiple game classes under environment-feedback coupling.

\subsection{Analytical Derivation of Transcritical Bifurcation}
Taking the temptation to defect $T$ as the core bifurcation control parameter and fixing payoff constants $R, S, P$, the replicator dynamic equation governing the evolution of cooperation frequency $x\in[0,1]$ is written as
\begin{equation}
\dot{x} = f(x;T) = x(1-x)(2x-1)\big[(S-P)+x(R-T-S+P)\big].
\label{eq:full_replicator}
\end{equation}
The fixed-point solutions are $x_1^*=0$, $x_2^*=1$, $x_3^*=1/2$, and the parameter-dependent interior equilibrium
\begin{equation}
x_4^*=\frac{P-S}{P-S+R-T}.
\end{equation}
For a generic transcritical bifurcation of one-dimensional system $\dot{x}=f(x;T)$ at critical point $(x^*,T_c)$, the full set of sufficient conditions are:
\begin{itemize}
\item Equilibrium condition: $f(x^*;T_c)=0$,
\item Linear degeneracy: $\dfrac{\partial f}{\partial x}\Big|_{(x^*,T_c)} = 0$,
\item Parameter critical condition: $f_T=\dfrac{\partial f}{\partial T}\Big|_{(x^*,T_c)} = 0$,
\item Non-degeneracy conditions:
\begin{itemize}
\item Transversality condition: $f_{xT}=\dfrac{\partial^2 f}{\partial x\partial T}\Big|_{(x^*,T_c)} \neq 0$,
\item Second-order non-degeneracy: $\dfrac{\partial^2 f}{\partial x^2}\Big|_{(x^*,T_c)} \neq 0$.
\end{itemize}
\end{itemize}

For our dynamics, the partial derivative with respect to bifurcation parameter $T$ reads
\begin{equation}
f_T=\frac{\partial f}{\partial T}=-x^2(1-x)(2x-1).
\end{equation}
The parameter critical condition $f_T=0$ restricts potential transcritical-bifurcation locations to $x=0$, $x=1/2$, and $x=1$, i.e., a transcritical bifurcation can only occur when $x_4^*$ collides with one of these three fixed points.
The mixed cross-partial derivative is
\begin{equation}
f_{xT}=\frac{\partial^2 f}{\partial x\partial T}=8x^3-9x^2+2x = x\left(8x^2-9x+2\right),
\end{equation}
whose roots are $x=0$, $x=\frac{9-\sqrt{17}}{16}\approx 0.305$, $x=\frac{9+\sqrt{17}}{16}\approx 0.820$.
Detailed polynomial algebraic expansions are omitted here and can be cross-checked via symbolic computation software.

Analytically solving the fixed-point collision condition $x_4^*=x^*$ yields two explicit critical parameter values for transcritical bifurcations, valid under the prerequisite $S\neq P$ to satisfy non-degeneracy:
\begin{equation}
T=R,\qquad T=R+S-P.
\end{equation}
These two values are sorted by magnitude for given payoff parameters and denoted as $T_{c1}$ and $T_{c2}$. Two distinct bifurcation events arise:
\begin{enumerate}
\item At $T=R$: $x_4^*$ collides with boundary equilibrium $x^*=1$ and they exchange stability. Crossing this threshold may move $x_4^*$ into or out of the feasible interval $(0,1)$.
\item At $T=R+S-P$: interior fixed points $x_3^*=1/2$ and $x_4^*$ coincide and interchange their stability, while boundary equilibria retain their stability properties.
\end{enumerate}
The sequential order of these two bifurcation events depends on the punishment parameter $P$. For our test‑case parameters $R=3$, $S=1$:
\begin{itemize}
\item $P=0$: $R=3 < R+S-P=4$, hence $T_{c1}=R=3$, $T_{c2}=R+S-P=4$;
\item $P=2$: $R+S-P=2 < R=3$, hence $T_{c1}=R+S-P=2$, $T_{c2}=R=3$.
\end{itemize}

For the special case $S=P$, $x_4^*=0$ coincides with $x^*=0$, but the transversality condition $f_{xT}|_{x^*=0}=0$ is violated; this corresponds to a higher-order degenerate scenario rather than a standard transcritical bifurcation.

Define perturbation $u=x-x^*$ and parameter offset $\mu=T-T_c$. A second-order truncated Taylor expansion yields the normal-form equation for transcritical bifurcation
\begin{equation}
\dot{u}=a\mu u -b u^2,
\label{eq:transcritical_normal}
\end{equation}
with bifurcation coefficients $a=f_{xT}|_{x^*}$, $b=-\frac12 f_{xx}|_{x^*}$, valid for $S\neq P$.

For one-dimensional continuous autonomous ODEs Hopf bifurcations cannot occur. Within the feasible parameter domain of this feedback-evolving game, all observed bifurcations satisfy the parameter critical condition $f_T=0$ and belong to transcritical type; saddle-node bifurcations do not arise.
The magnitude‑ordered critical thresholds $T_{c1}$ and $T_{c2}$ from analytical expressions are cross-validated against numerically computed bifurcation diagrams.

The above theoretical work derives explicit stability criteria for all fixed points and identifies critical bifurcation thresholds modulated by payoff parameters. It also elaborates the sequential transition rule of game regimes driven by payoff parameters. To visually validate these analytical results and further excavate practical regulatory insights for cooperative behaviors, Section 4 conducts multi-faceted numerical simulations, including global parameter scanning, single-parameter bifurcation tracing, local phase-plane verification, and extended simulations for spatially structured populations.

\section{Results and discussion}
\label{section.4}
This section employs numerical simulations to systematically verify the previous theoretical analysis and dynamic predictions. Combined with global phase diagrams, single-parameter dynamic evolution analysis, and phase-plane evolutionary trajectories, this section reveals the regulatory mechanisms of core payoff parameters on the steady states and evolutionary trajectories of cooperative strategies.

\subsection{Joint Regulation of Initial Cooperation Frequency and Payoff Parameters}
Figure~\ref{fig.1} shows the distribution of steady-state cooperation frequencies generated by system evolution on the $T-P$ parameter plane under different initial cooperation frequencies $x_0$. The horizontal axis represents the temptation to defect, with $T\in[0,6]$, and the vertical axis denotes the punishment payoff $P\in[-1,3]$. The color gradient characterizes the final steady-state cooperation frequency $x^*$ after evolution. According to the critical boundaries $T=3$ and $P=1$, the parameter plane is divided into four typical game regions: the region $T<3$ and $P>1$ corresponds to SH; $T>3$ and $P>1$ corresponds to PD; $T<3$ and $P<1$ corresponds to HG; and $T>3$ and $P<1$ corresponds to SD. Evolutionary outcomes under different initial conditions exhibit strong parameter dependence and initial-condition sensitivity:

\begin{figure}
  \centering
  \includegraphics[width=1.0\textwidth,clip=false]{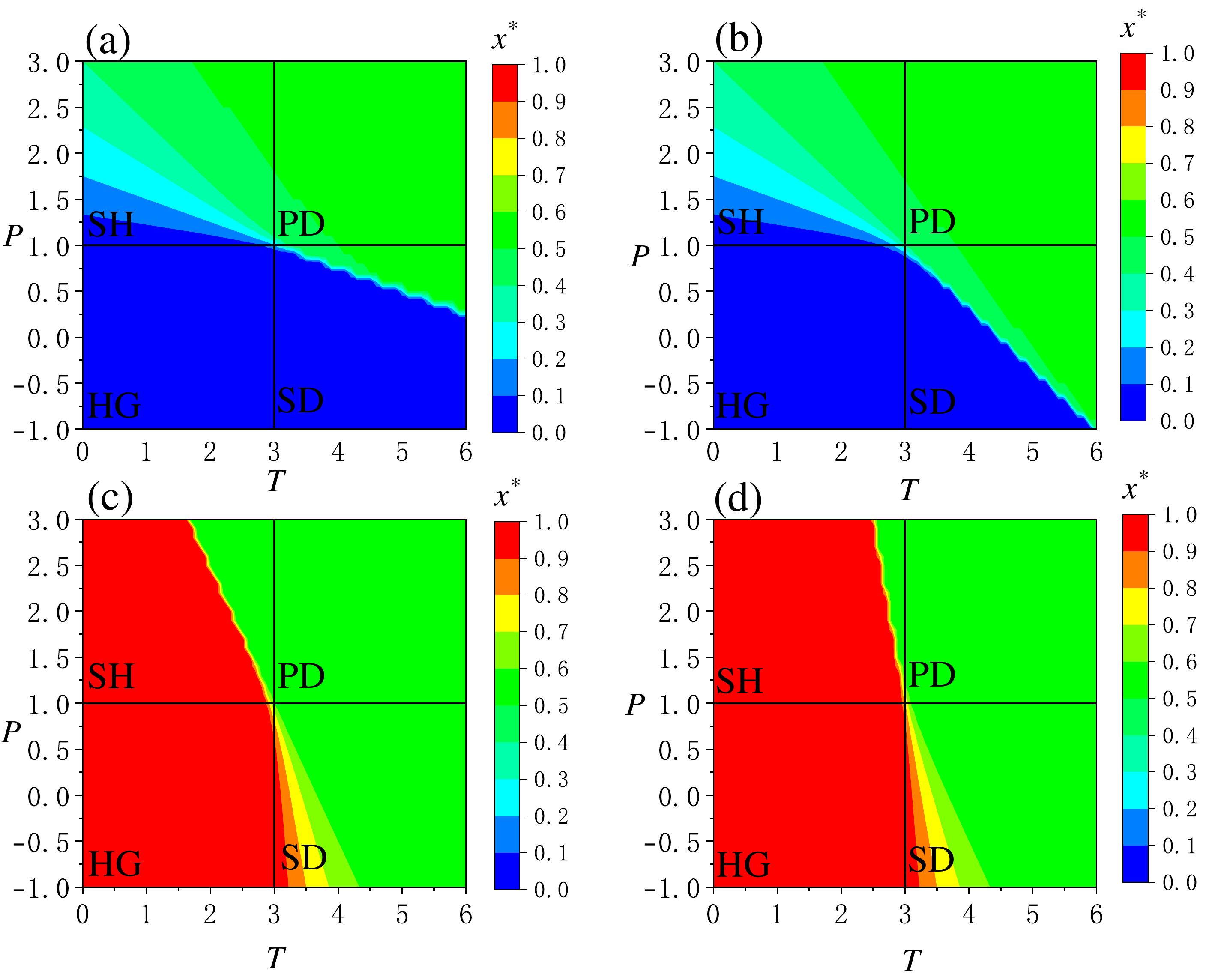}
  \caption{Steady-state cooperation frequency $x^*$ on the $T-P$ parameter plane under different initial cooperation levels $x_0$.
  (a) Heat map of $x^*$ for low initial cooperation $x_0=0.2$.
  (b) Heat map of $x^*$ for medium-low initial cooperation $x_0=0.4$.
  (c) Heat map of $x^*$ for medium-high initial cooperation $x_0=0.6$.
  (d) Heat map of $x^*$ for high initial cooperation $x_0=0.8$.
  The horizontal axis denotes the temptation to defect $T\in[0,6]$, and the vertical axis represents the punishment payoff $P\in[-1,3]$.
  The color scale represents $x^*$ ranging from 0 (full defection, blue) to 1 (full cooperation, red).
  The parameter plane is divided into four classic game regions by $T=3$ and $P=1$: Stag-Hunt (SH), Prisoner's Dilemma (PD), Harmony Game (HG), and Snowdrift Game (SD).
  Simulation results are averaged over 30 independent runs.}
  \label{fig.1}
\end{figure}

(a) $x_0 = 0.2$ (Low initial cooperation level)

For low initial cooperation $x_0=0.2$, the population exhibits weak cooperative norms and lacks sufficient altruistic foundations. Within the SH regime, equilibrium cooperation rises as $P$ increases and $T$ decreases, which indicates that stringent punishment paired with low temptation constitutes the only viable approach to promote cooperation for such populations. The PD parameter domain uniformly converges to $x^*=1/2$, which verifies the strong robustness of its mixed-strategy equilibrium. In the HG regime, full defection prevails across most parameter space; stable universal full cooperation is only achievable within a narrow parameter interval featuring high punishment and low temptation. This observation confirms the pronounced initial-condition sensitivity of bistable game systems as well as the inherent robustness of the PD mixed equilibrium. It further reveals that early stringent punishment is essential to curb widespread free-riding in communities lacking spontaneous cooperative tendencies.

(b) $x_0 = 0.4$ (medium–low initial cooperation level)

The evolutionary trends under $x_0=0.4$ are broadly consistent with those observed for $x_0=0.2$. Within the SH regime, the gradient of equilibrium cooperation becomes gentler, and the stationary cooperation level rises marginally under identical parameter sets. This illustrates that a moderate increase in initial cooperation facilitates prosocial behaviors, yet high $P$ combined with low $T$ remains indispensable to sustain stable cooperation. The PD domain always converges to $x^*=1/2$, displaying strong robustness with respect to initial values. In the SH regime, the system converges to either the full-cooperation or all-defection boundary fixed points with no interior equilibrium present, whereas evolutionary trajectories within the SD regime are governed by initial strategies under simplified dynamical rules. The system retains relatively stable cooperation when $x_0$ lies in the range $[0.2,0.4]$, while it evolves toward universal full defection or global full cooperation once $x_0$ moves outside this interval. It can be concluded that the initial cooperation fraction dominates long-run evolutionary outcomes, and the parameter-dependent interior fixed point serves as the core factor determining overall system stability.

(c) $x_0 = 0.6$ (medium to high initial cooperation level)

Once initial cooperation exceeds the critical threshold, the system’s dynamical behaviors alter markedly. The SH regime almost always converges to full cooperation $x^*=1$; convergence toward the mixed equilibrium $x^*=1/2$ only emerges under jointly high $T$ and high $P$. The PD domain reliably settles at the mixed equilibrium $x^*=1/2$ and exhibits negligible sensitivity to initial conditions. For the HG regime, high initial cooperation renders full cooperation $x^*=1$ the dominant attracting steady state. In the SD regime, equilibrium cooperation declines distinctly as $T$ increases and $P$ decreases, with defection acting as the persistent long-run outcome and global full cooperation unattainable. Cooperation peaks around the critical boundaries $T=3$ and $P=1$. Larger values of $T$ paired with smaller $P$ yield low cooperation levels, and no parameter combination within this regime can achieve universal full cooperation. This finding demonstrates that a favorable initial cooperative atmosphere can effectively enlarge the basin of attraction for full cooperation in bistable game regimes.

(d) $x_0 = 0.8$ (high initial cooperation level)

A high initial cooperation fraction creates favorable prerequisites for sustained collective cooperation. Within the SH regime, most parameter combinations converge to full cooperation, and only narrow domains with simultaneously high $T$ and high $P$ yield the mixed equilibrium. The PD domain always stabilizes at the mixed-strategy equilibrium, which is robust to both initial states and external perturbations. All parameter sets in the HG regime converge to full cooperation, confirming cooperation as the dominant evolutionary trend. For the SD regime, long-run outcomes depend jointly on initial cooperation levels and payoff parameters: low temptation combined with stringent punishment pushes trajectories toward full cooperation, while high $T$ paired with low $P$ leads to universal full defection. This observation reveals that cultivating a high initial cooperation level can substantially stabilize collective cooperative behaviors under regular governance.

Cross-comparison across all initial cooperation fractions demonstrates that long-run cooperative steady states are jointly determined by initial cooperation levels and payoff parameters. The SH and HG regimes exhibit prominent initial-condition sensitivity, wherein a higher initial cooperation fraction facilitates the formation and persistence of full cooperation. In contrast, the mixed equilibrium within the PD domain presents strong robustness and is independent of initial values. Defection always prevails throughout the SD regime; cooperation can only persist within narrow parameter intervals, and global full cooperation is unattainable. The core parameters $T$ and $P$ define critical thresholds governing evolutionary tendencies: low temptation paired with stringent punishment favors cooperation, whereas initial cooperation acts as a vital threshold controlling equilibrium outcomes in bistable regions. Collectively, these factors form an integrated global dynamical framework describing the evolution of cooperation.

Based on the parameter boundaries extracted from the global phase diagram in Figure~\ref{fig.1}, Figure~\ref{fig.2} and~\ref{fig.3} further illustrate evolutionary trajectories to quantify how temptation to defect $T$ modulates system steady states. With punishment fixed at $P=0$ and $P=2$, the two figures show continuous variations in the existence and stability of fixed points as $T$ varies, together with regime transitions across distinct game types. This elaborates the switching mechanism of cooperative dynamics from a single-parameter perspective.

\subsection{Dynamic Evolution Driven by Temptation to Defect and Continuous Transition of Game Types}
\begin{figure}
  \centering
  \includegraphics[width=0.85\textwidth,clip=false]{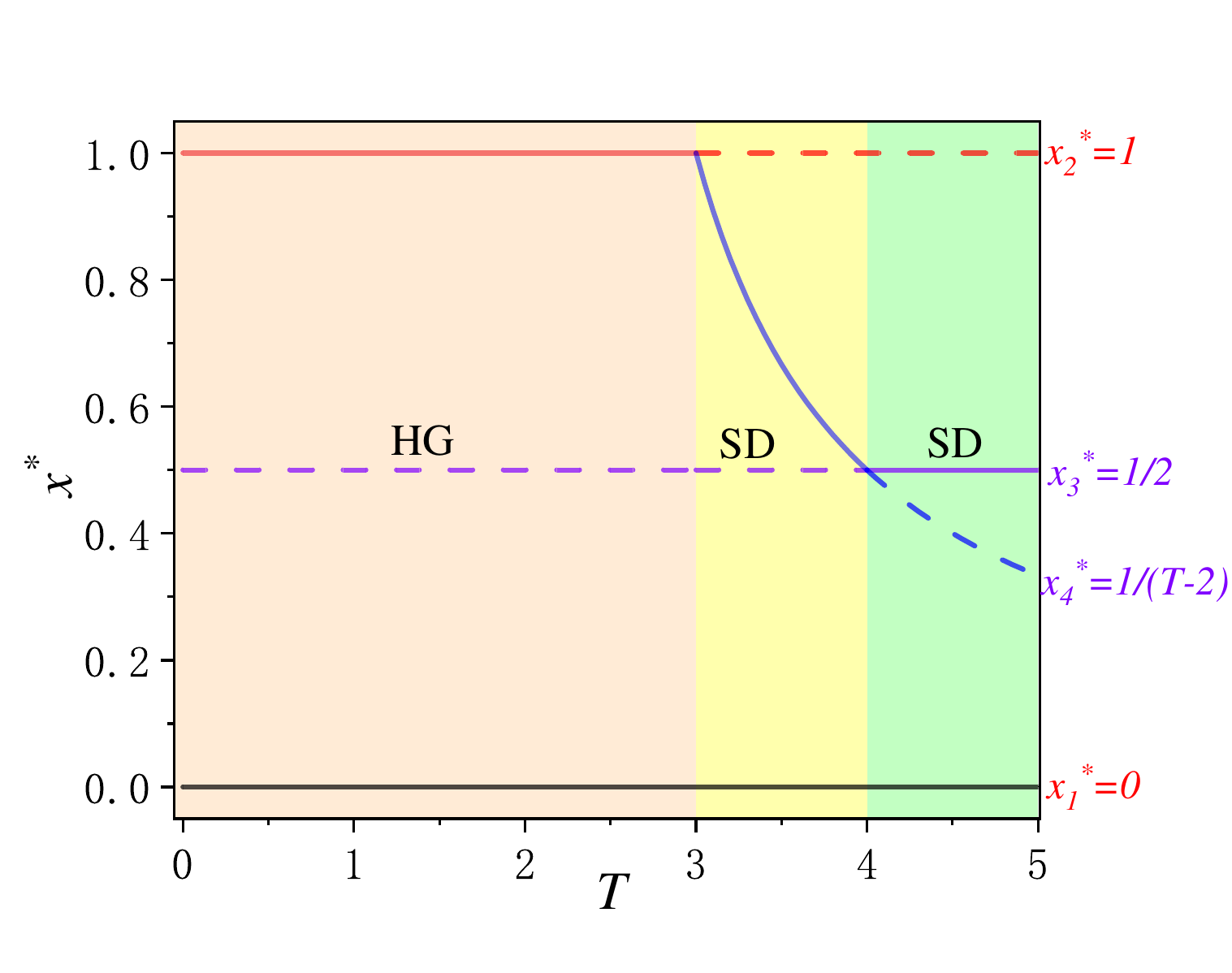}
  \caption{Bifurcation diagram of the replicator dynamic system with fixed punishment payoff $P=0$.
  The horizontal axis is the temptation to defect $T\in[0,5]$, and the vertical axis denotes the long-run steady-state cooperation fraction $x^*$. Solid lines mark locally asymptotically stable fixed points, and dashed lines mark unstable ones. Based on game types and stability features, the parameter space is divided into three dynamical regimes: (i) bistable Harmony Game (HG), $T\in[0,T_{c1}]$; (ii) transitional Snowdrift Game (SD), $T\in(T_{c1},T_{c2}]$; (iii) Snowdrift Game (SD), $T>T_{c2}$. Convergence behavior of the system strongly depends on initial cooperation fractions within both HG and SD regimes.
  For $T\in[0,T_{c1}]$, the unstable interior fixed point $x_3^*=1/2$ serves as a separatrix: trajectories with $x_0>1/2$ converge to full cooperation $x_2^*=1$, while those with $x_0<1/2$ converge to full defection $x_1^*=0$. At the HG–SD boundary $T=T_{c1}$, stability exchange takes place among interior fixed points. In the SD parameter domain, the system exhibits bistability: the locally stable all-defection equilibrium $x_1^*=0$ coexists with a locally stable interior mixed-strategy equilibrium. The unstable interior fixed point acts as separatrix between their basins of attraction. When $T>T_{c2}$, stability exchange occurs between the two interior fixed points $x_3^*$ and $x_4^*$, and $x_4^*$ remains within the feasible interval $[0,1]$. The system remains within the SD regime and still shows sensitivity to initial conditions.
  For this specific parameter set ($R=3$, $S=1$, $P=0$), we have $T_{c1}=3$ and $T_{c2}=4$. The numerically observed bifurcation thresholds are cross-validated against the analytical thresholds for transcritical bifurcations derived in Section 3.6.}
  \label{fig.2}
\end{figure}

Figure~\ref{fig.2} plots the $T-x$ bifurcation curve for $P=0$ under fixed payoff parameters $R=3$ and $S=1$, illustrating how temptation $T$ drives global evolutionary trends and triggers regime shifts between different dynamical phases.
For $T\in[0,T_{c1}]$, the system lies in the bistable Harmony Game (HG) regime with two stable boundary equilibria. The final equilibrium solely depends on initial $x_0$: trajectories starting at $x_0\geq 1/2$ converge to full cooperation $x^*=1$, while those with $x_0<1/2$ converge to full defection $x^*=0$, showing obvious path dependence on initial states. $T_{c1}=3$ is the first transcritical-bifurcation critical point for this parameter setup. At this threshold, the full-cooperation equilibrium loses stability, and stability interchange occurs among interior fixed points, transferring the system from HG to the Snowdrift Game (SD). For $T\in(T_{c1},T_{c2})$, the system exhibits bistability within the SD regime: the locally stable full-defection equilibrium $x^*=0$ coexists with a locally stable interior mixed-strategy equilibrium. An unstable interior fixed point acts as separatrix, and evolutionary outcomes remain sensitive to initial conditions. The second transcritical-bifurcation critical point emerges at $T_{c2}=4$. At this threshold, stability exchange takes place between the two interior fixed points $x_3^*$ and $x_4^*$; the parameter-dependent interior equilibrium $x_4^*$ stays inside the feasible interval, and the system remains within the Snowdrift Game regime. For $T>T_{c2}$, bistable characteristics persist and evolutionary trajectories still depend on initial cooperation fractions. Excessive temptation suppresses cooperative payoffs, but the system still maintains mixed-strategy coexistence rather than collapsing to full defection. This bifurcation curve confirms that $T$ dominates the overall evolutionary direction of group cooperation.

\begin{figure}
  \centering
  \includegraphics[width=0.85\textwidth,clip=false]{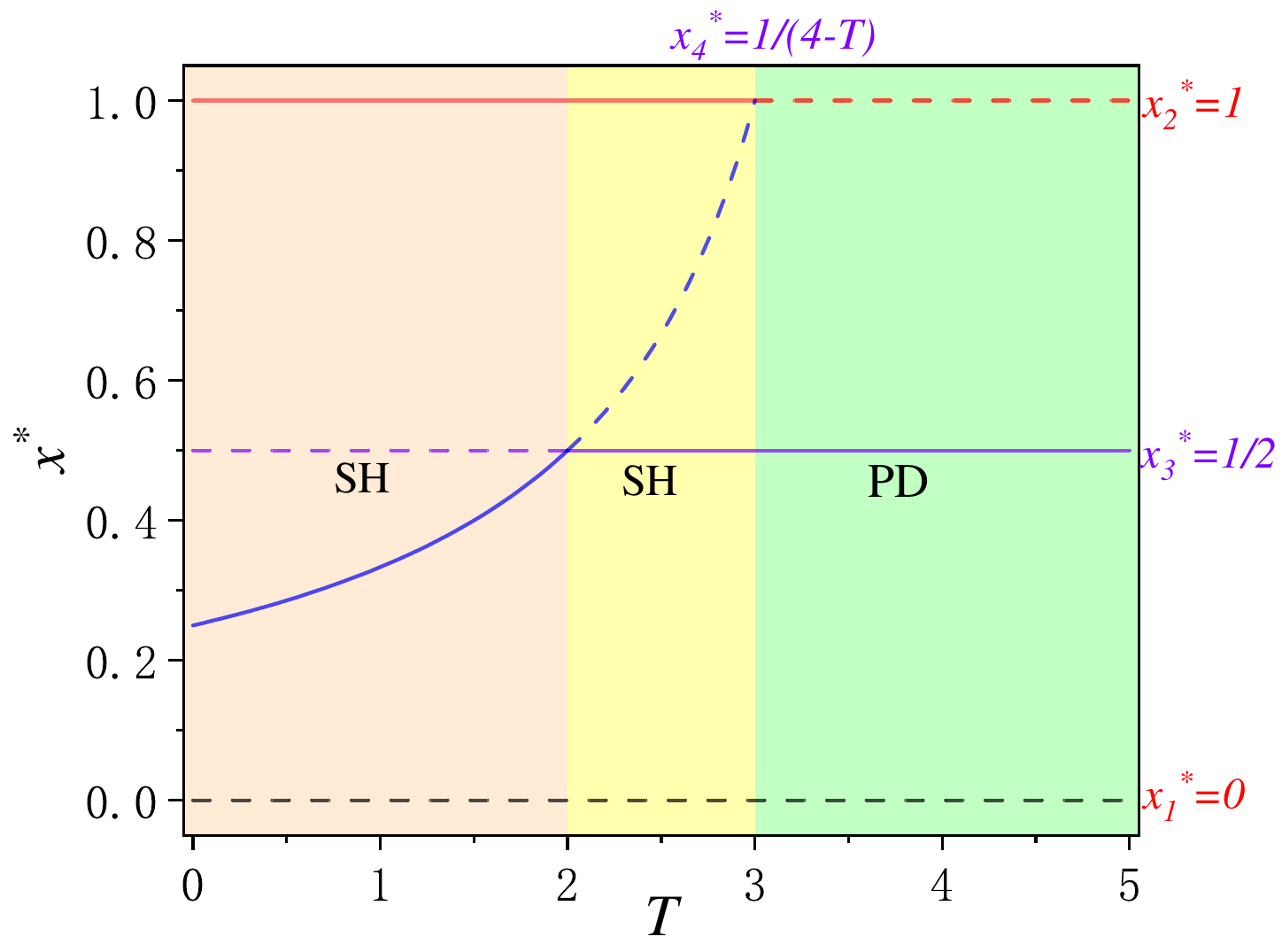}
  \caption{Bifurcation diagram of the replicator dynamic system with fixed punishment payoff $P=2$.
  The horizontal axis is the temptation to defect $T\in[0,5]$, and the vertical axis stands for the long-run steady-state cooperation fraction $x^*$. Solid lines mark locally asymptotically stable fixed points, and dashed lines mark unstable ones. The parameter space is split into three dynamical regimes according to game types and stability properties: (i) Stag-Hunt (SH) with interior equilibrium, $T\in[0,T_{c1}]$; (ii) transitional SH with stability exchange, $T\in(T_{c1},T_{c2}]$; (iii) Prisoner's Dilemma (PD) with a unique mixed-strategy steady state, $T>T_{c2}$. Convergence behavior of the system strongly depends on initial cooperation fractions within the SH regimes.
  For $T\in[0,T_{c1}]$, both the interior equilibrium $x_4^*=1/(4-T)$ and full-cooperation equilibrium $x_2^*=1$ are stable, with the unstable point $x_3^*=1/2$ acting as a separatrix. Trajectories with $x_0<1/2$ converge to $x_4^*=1/(4-T)$, while those with $x_0>1/2$ converge to full cooperation $x_2^*=1$. At $T=T_{c1}$, a stability exchange induced by transcritical bifurcation takes place between $x_3^*=1/2$ and $x_4^*$. Within $T\in(T_{c1},T_{c2}]$, $x_3^*=1/2$ becomes stable and $x_4^*$ turns unstable; trajectories with $x_0<1/2$ converge to the mixed equilibrium $x_3^*=1/2$, and $x_0>1/2$ still converge to full cooperation.
  When $T>T_{c2}$, the system enters the PD regime. The full-cooperation equilibrium $x_2^*=1$ loses stability, $x_4^*$ exits the feasible interval, and only $x_3^*=1/2$ acts as the unique locally stable interior equilibrium. All trajectories inside $(0,1)$ converge to $x_3^*=1/2$, revealing the intrinsic social dilemma of PD.
  For this specific parameter set ($R=3$, $S=1$, $P=2$), we have $T_{c1}=2$ and $T_{c2}=3$. The numerically obtained bifurcation thresholds are cross-validated with analytically derived thresholds for transcritical bifurcations from Section 3.6.}
  \label{fig.3}
\end{figure}

Figure~\ref{fig.3} displays outcomes under higher punishment $P=2$, with payoff parameters fixed at $R=3$ and $S=1$. For $T\in[0,T_{c1}]$, the system falls within the Stag-Hunt (SH) regime: $x^*=0$ is unstable, $x^*=1$ is a locally stable attractor, $x^*=1/2$ acts as an unstable separatrix, and the interior equilibrium $x_4^*$ maintains local stability within $x\in(0,1)$. $T_{c1}=2$ is the first transcritical-bifurcation critical threshold for stability exchange between $x^*=1/2$ and $x_4^*$: $x^*=1/2$ turns stable whereas $x_4^*$ loses stability, yet the system remains in the SH regime for $T\in(T_{c1},T_{c2}]$. At $T_{c2}=3$, the second transcritical-bifurcation threshold is reached. Once $T>T_{c2}$, the system transitions to PD. The two boundary equilibria $x^*=0$ and $x^*=1$ both turn unstable, $x_4^*$ exits the feasible interval and disappears, and only $x^*=1/2$ survives as the unique locally stable interior equilibrium. The system eventually stabilizes at a fixed mixed state, which precisely reflects the fundamental conflict between individual self-interest and collective welfare inherent to the Prisoner’s Dilemma under environmental-feedback coupling.

Comparing Figure~\ref{fig.2} ($P=0$) and Figure \ref{fig.3} ($P=2$) reveals how punishment payoff $P$ remodels global dynamical landscapes. For our fixed payoff setting $R=3$, $S=1$, while both plots take $T$ as the main bifurcation parameter and possess two critical transition thresholds $T_{c1}$ and $T_{c2}$, they differ greatly in critical values, sequence of game regime shifts and equilibrium structures. Under low punishment $P=0$, for $R=3,S=1$, transcritical bifurcations arise at $T_{c1}=3$ and $T_{c2}=4$, switching the system from HG to SD. Bistability persists for $T>3$ within the SD regime, and the system can converge to mixed-strategy states or full defection depending on initial conditions. Under high punishment $P=2$, transcritical bifurcation thresholds move left to $T_{c1}=2$ and $T_{c2}=3$, leading to a direct transition from SH to PD; full cooperation stays stable for all $T<T_{c2}$.Overall, larger punishment payoffs strengthen the stability of cooperative equilibria and shift the critical temptation threshold at which cooperation collapses to a lower value. This confirms that payoff structures act as core regulators of evolutionary trajectories and long-run steady states of cooperation.

\subsection{Verification of Local Dynamics and Global Evolution}
\begin{figure}
  \centering
  \includegraphics[width=0.9\textwidth,clip=false]{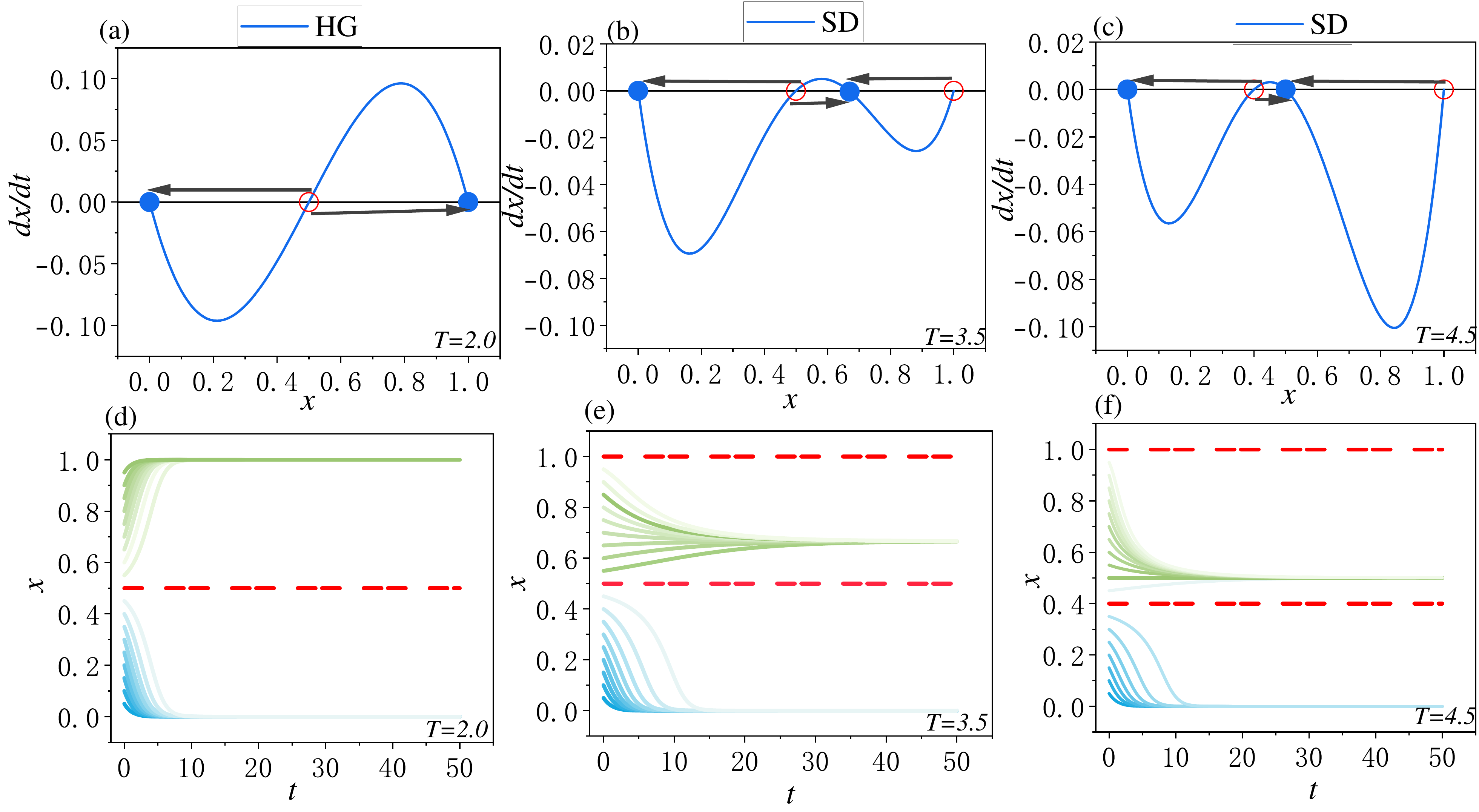}
  \caption{Phase-plane portraits of $\mathrm{d}x/\mathrm{d}t$ versus $x$ (top row) and corresponding time-evolution trajectories of cooperation frequency $x(t)$ (bottom row). Subplots correspond to different game regimes: (a) HG with $T=2.0$, (b) SD with $T=3.5$, (c) SD with $T=4.5$; panels (d), (e), (f) are the time-series curves matching (a), (b), (c) respectively. Curve intersections with the horizontal axis represent evolutionary fixed points, and arrows indicate the direction of strategy evolution. Trajectory colors differentiate initial cooperation fractions: blue for $x_0<1/2$, red dashed line for the unstable separatrix $x_0=1/2$, green for $x_0>1/2$.}
  \label{fig.4}
\end{figure}

This subsection combines phase-plane analysis and time-series trajectories to verify the theoretically derived fixed-point stability and evolutionary rules across all game regimes. Figure \ref{fig.4} displays phase portraits in the upper row and corresponding time-evolution curves below. For each phase portrait, the horizontal axis is the cooperation fraction $x$, and the vertical axis represents the evolutionary rate $\mathrm{d}x/\mathrm{d}t$. Intersections between curves and the horizontal axis identify fixed points and their stability, while arrows indicate evolutionary directions to illustrate dynamical flows under distinct initial $x_0$. The bottom row plots cooperation fraction $x(t)$ versus evolutionary time $t$. Simulations with multiple initial values are highly consistent with phase-plane results, which strongly verify the global attraction properties of all equilibria.

For the SD case with $T=3.5$, $x_3^*=1/2$ serves as an unstable separatrix between full defection and the stable interior equilibrium, and time-series curves clearly show initial-value-dependent convergence. At $T=4.5$, stability exchange occurs between interior fixed points: the separatrix shifts from $x_3^*$ to $x_4^*$, and all trajectories converge to the mixed steady state, confirming structural transformations near bifurcation thresholds. The HG regime for $T<3$ exhibits standard bistable dynamics, where the saddle point $x_3^*=1/2$ strictly splits two basins of attraction; time trajectories further demonstrate that initial cooperation governs equilibrium selection. All observations agree with previous theoretical deductions on regime transition mechanisms. Joint evidence from phase portraits and time series verifies core features including HG bistability and interior stability reversal in SD. This cross validation achieves perfect consistency between local dynamical mechanisms and macroscopic evolution, providing robust dynamical evidence for the regulation rules of cooperative steady states.

\begin{figure}
  \centering
  \includegraphics[width=0.9\textwidth,clip=false]{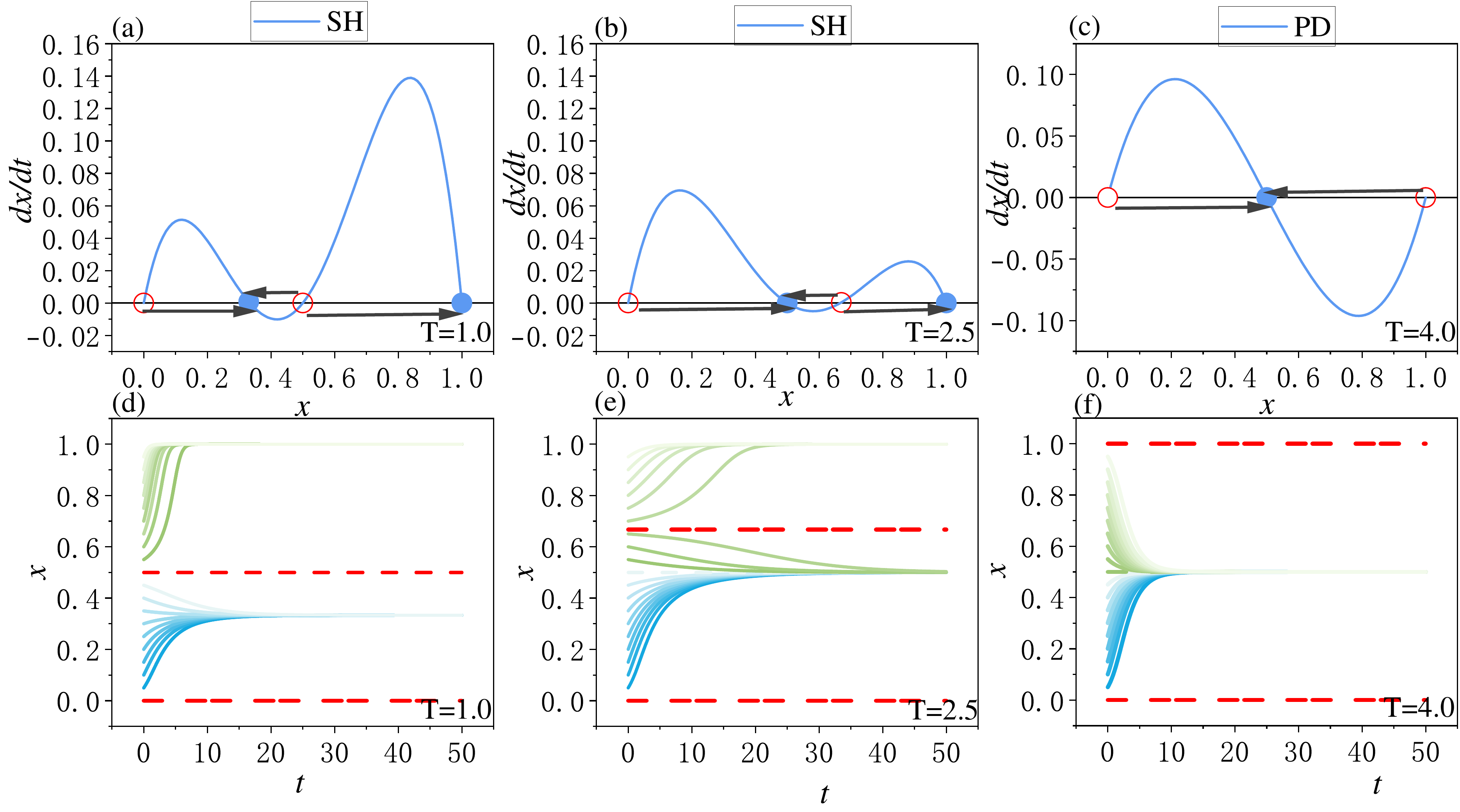}
  \caption{Phase-plane portraits of $\mathrm{d}x/\mathrm{d}t$ versus $x$ (top row) and corresponding time-evolution trajectories of cooperation frequency $x(t)$ (bottom row). Subplots cover two SH cases ($T=1.0$, $T=2.5$) and one PD case ($T=4.0$), where panels (d), (e), (f) correspond to (a), (b), (c) in sequence. Fixed points lie on horizontal-axis intersections, and evolutionary directions are marked by arrows. The color coding rule for initial cooperation fractions follows the definition in Figure 4.}
  \label{fig.5}
\end{figure}

Figure \ref{fig.5} verifies the local dynamics and evolutionary trajectories under $P=2$. The graphical settings are consistent with those in Figure \ref{fig.4}, and simulation results under multiple initial conditions agree well with the phase-plane conclusions.In the SH regime with $T=1.0$ and $T=2.5$, the results demonstrate the threshold effect induced by stability transitions of interior fixed points and exhibit typical initial-condition dependence. In the PD regime for $T>3$, all boundary steady states lose stability, and the system globally converges to a unique mixed steady state. These observations fully validate the structural transition of steady-state configurations during the evolutionary process from the SH regime to the PD regime, achieving closed-loop self-consistency between local dynamic analysis and global evolutionary behaviors.

\subsection{Effects of Initial Spatial Distributions on Steady-State Results}
To uncover how spatial structure regulates steady states and dynamical behaviors of cooperative evolution, we perform agent-based evolutionary game simulations on a 2D square lattice with periodic boundaries. Each agent interacts with its four nearest neighbors and updates strategies via an imitation-update process adopting the Fermi-form logit function. We systematically plot phase diagrams of steady-state cooperation fraction $x^*$ against temptation to defect $T$, together with time-series trajectories under typical parameters, to clarify the key regulatory impacts of spatial topology on cooperative equilibria, bifurcations and dynamical features.

\begin{table}[htbp]
\centering
\renewcommand{\arraystretch}{1.35}
\setlength{\tabcolsep}{8pt}
\caption{Comparison between well-mixed mean-field replicator dynamics and spatial lattice evolutionary game model.}
\label{tab:model_comparison}
\begin{tabularx}{\linewidth}{l >{\raggedright\arraybackslash}X >{\raggedright\arraybackslash}X}
\hline
Feature & Well-mixed mean-field model & Spatial lattice model \\
\hline
Spatial structure & Fully‑connected, structure-free population & 2D square lattice with von Neumann four-neighbor topology \\
\hline
Strategy update & Deterministic replicator evolution & Stochastic Fermi-logit imitation \\
\hline
Selection noise & Implicitly embedded in payoff differences & Explicit noise coefficient $w=\beta^{-1}=0.01$ \\
\hline
Dynamic nature & Continuous deterministic dynamics with analytical fixed points & Discrete stochastic Monte-Carlo dynamics with statistical steady states \\
\hline
Core evolutionary mechanism & Global frequency-dependent selection & Local interactions and spatial reciprocity \\
\hline
\multirow{4}{*}{\centering Equilibrium behaviour}
& HG: Bistable & HG: Full-cooperation dominance \\
\cline{2-3}
& SH: Bistable & SH: Full-cooperation dominance \\
\cline{2-3}
& SD: Bistable (all-defection plus a mixed state) & SD: Low-cooperation \\
\cline{2-3}
& PD: mixed-strategy equilibrium & PD: Low-cooperation \\
\hline
\end{tabularx}
\end{table}

\begin{figure}
  \centering
  \includegraphics[width=0.9\textwidth]{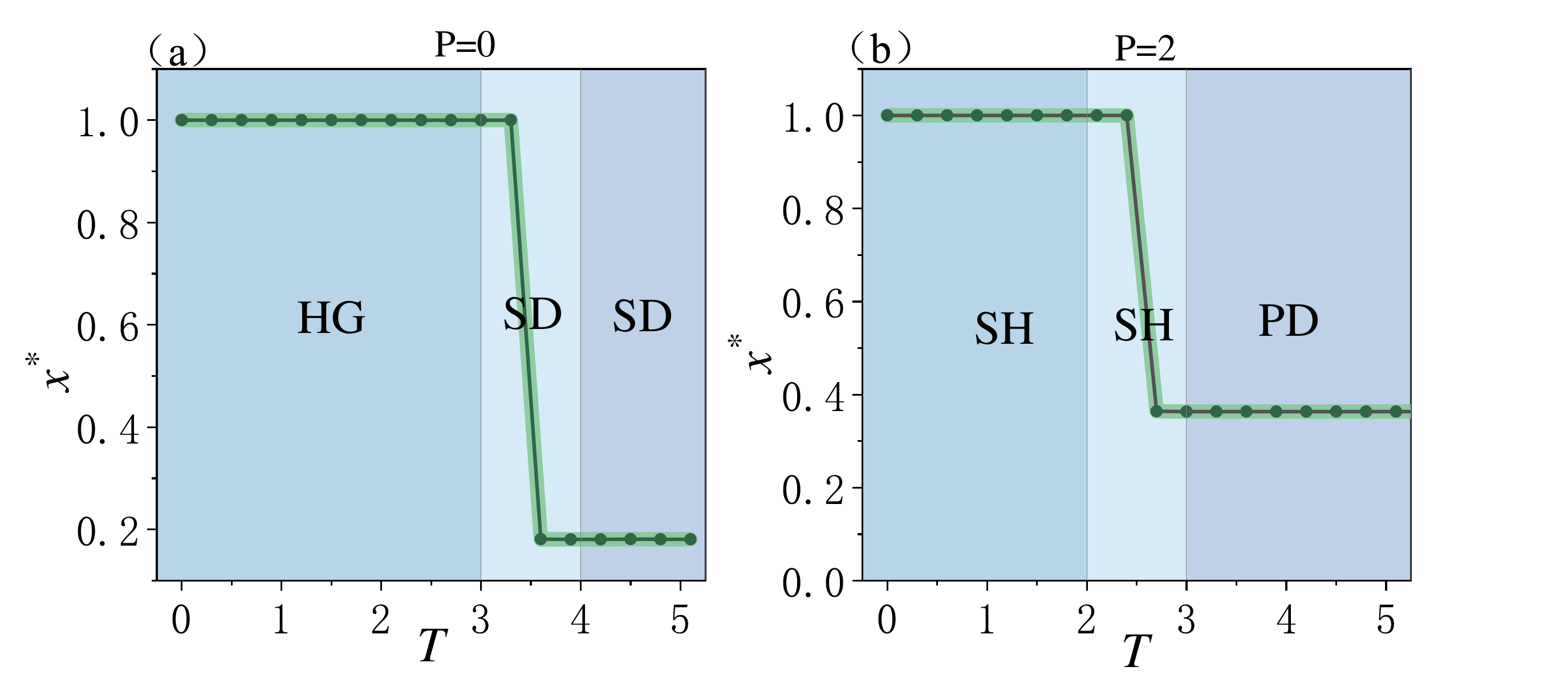}
  \caption{Steady-state cooperation frequency $x^*$ versus defection temptation $T$ for the two-dimensional square lattice evolutionary game system. The simulation adopts a lattice size $L=100$ ($N=L^2$), each individual interacts with 4 nearest neighbors under periodic boundary conditions, and the noise coefficient of the Fermi-form logit imitation kernel is set as $w=\beta^{-1}=0.01$. The vertical axis represents the steady‑state cooperation frequency $x^*$. The parameter range is divided into dynamical regimes based on game type and stability behavior:
  (a) $P=0$: (i) HG regime ($T\in[0,3]$) with globally stable full cooperation; (ii) SD regime ($T>3$) with a stable non-zero low-cooperation steady state.
  (b) $P=2$: (i) SH regime ($T\in[0,2]$) where full cooperation is globally attractive; (ii) transitional regime ($T\in(2,3]$); (iii) PD regime ($T>3$) with a stable mixed-strategy steady state as the unique global attractor.
  Spatial lattice structure eliminates the full-defection absorbing state in both cases. All steady outcomes are averaged over 50 independent realizations, with $10^5$ thermalization Monte Carlo steps and $5\times10^4$ sampling steps for each realization.}
  \label{fig.6}
\end{figure}

Figure \ref{fig.6} plots steady-state cooperation fraction $x^*$ versus $T$ for the lattice system under two punishment values $P=0$ and $P=2$. Color-coded areas represent different dynamical regimes, intuitively showing phase transitions and stationary properties of cooperation under spatial interactions. Unlike well-mixed populations, lattice structures rule out the full-defection absorbing state via spatial cooperative clusters, sustaining persistent non-zero cooperation. For $P=0$, the system stays in the HG phase over $T\in[0,3]$, where global full cooperation $x^*=1$ is dominant. Once $T$ exceeds the critical threshold $T\approx3$, the system shifts to the SD phase; the stationary cooperation level drops sharply and stabilizes at a low non-zero value $x^*\approx0.18$.
At higher punishment $P=2$, the system presents qualitatively different phase structures. For $T\in[0,2]$, the lattice lies in the SH phase with globally attractive full cooperation. When $T>3$, the system enters the PD phase and stabilizes at $x^*\approx0.35$, a value obviously higher than the stationary cooperation fraction under $P=0$. Comparative analysis indicates that spatial topology eliminates the full-defection absorbing state in both parameter cases and maintains cooperation by forming cooperative clusters. Larger $P$ substantially improves long-run cooperation levels, reshapes phase boundaries and shifts critical transition thresholds. These findings prove that spatial structure and payoff parameters jointly determine the global dynamical features of cooperative evolution.

\begin{figure}
  \centering
  \includegraphics[width=0.9\textwidth,clip=false]{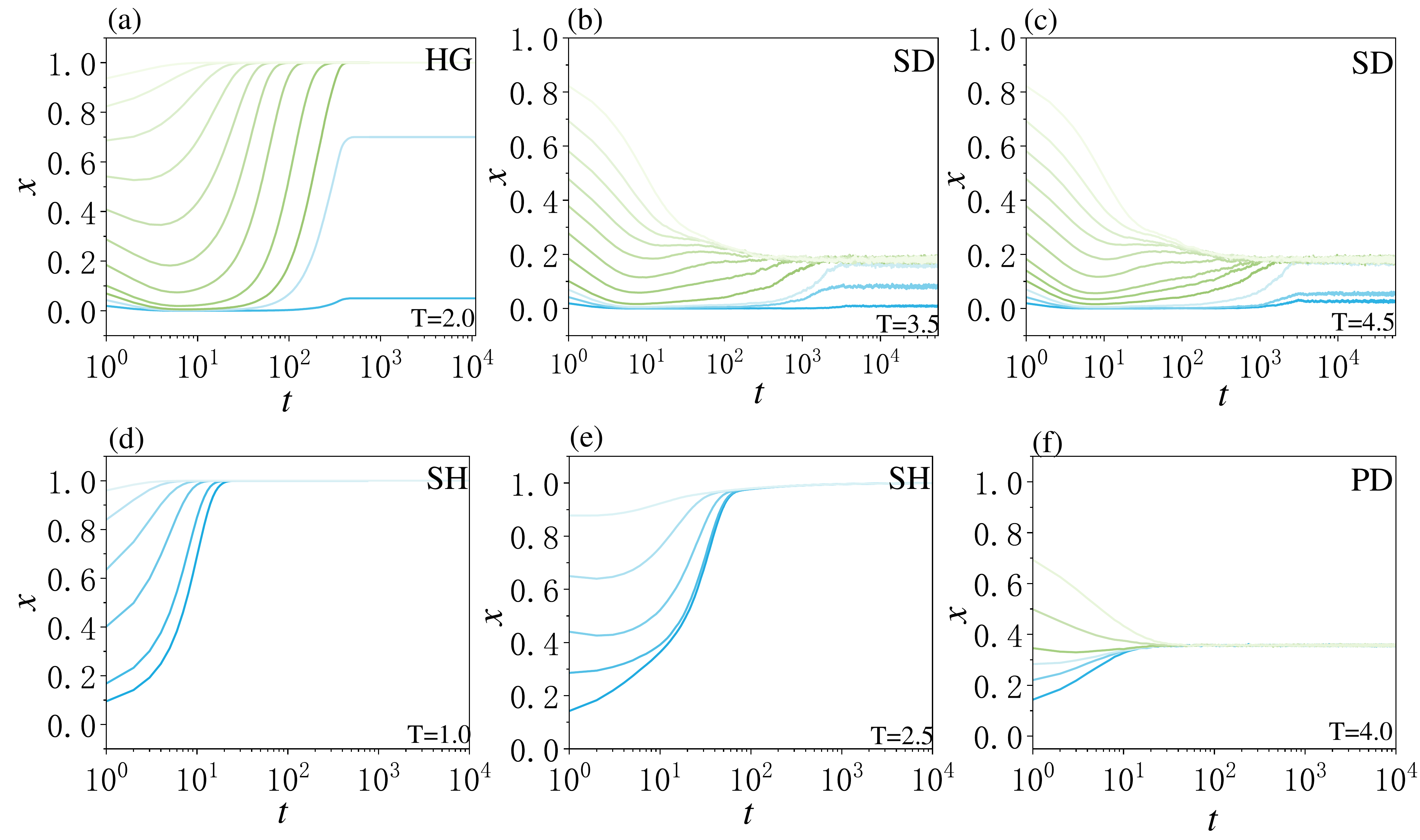}
  \caption{Time-evolution trajectories of cooperation fraction $x(t)$ on the 2D lattice.
  Panels (a)--(c) show results for $P=0$ at three typical temptation values:
  (a) HG, $T=2.0$; (b) SD, $T=3.5$; (c) SD, $T=4.5$.
  Panels (d)--(f) show results for $P=2$:
  (d) SH, $T=1.0$; (e) SH, $T=2.5$; (f) PD, $T=4.0$.
  The horizontal axis is evolutionary time $t$, and the vertical axis denotes cooperation fraction $x(t)$.
  All trajectories converge to full cooperation $x^*=1$ in HG and SH regimes.
  In SD, trajectories reach stable non-zero fluctuating equilibria that rely on initial values.
  In PD, all initial states converge to a single stable mixed equilibrium.
  Spatial topology eliminates the full-defection absorbing state and sustains non-zero cooperation.
  Each curve is averaged across 20 independent simulation trials.}
  \label{fig.7}
\end{figure}

Figure \ref{fig.7} shows the time evolution of cooperation fraction $x(t)$ on the 2D lattice under different game parameters, which clearly illustrates relaxation dynamics and stationary properties controlled by spatial topology and payoff parameters.
For $P=0$, rising temptation $T$ drives the system sequentially through HG and SD phases.
At $T=2.0$ (HG), full cooperation dominates globally; all trajectories starting above $x_0=0.1$ quickly converge to $x^*=1$.
At $T=3.5$ (SD), obvious bistability emerges: trajectories with initial fractions above roughly $0.15$ settle near $x\approx0.20$, while those below this threshold fall into a low-cooperation attractor and stabilize around $x\approx0.04$.
At $T=4.5$ (SD), stronger temptation suppresses cooperation overall. The stationary cooperation level rises with initial values and converges to roughly $x\approx0.15$, whereas trajectories starting from low initial fractions stabilize at an even lower $x\approx0.02$.

For $P=2$, the system exhibits distinct evolutionary dynamics across SH and PD phases.
At $T=1.0$ and $T=2.5$ (SH), full cooperation acts as the globally stable attractor; all initial states converge to $x^*=1$ with different relaxation times.
At $T=4.0$ (PD), the system settles into a fluctuating mixed equilibrium, where all trajectories fall within the same basin of attraction and stabilize near $x\approx0.35$.
These observations verify that spatial topology fundamentally removes the full-defection absorbing state. Raising punishment $P$ greatly improves cooperative stability and lifts long-run cooperation levels, and the two factors together determine the overall dynamical behaviors of cooperation evolution.

To examine finite-size effects, we supplemented control simulations with larger lattice sizes ($L=50,100,200$), and these convergence results are provided in the Response to Reviewers for the spatial lattice simulations shown in Figure 7. Under identical parameters and initial conditions, the stationary cooperation levels and qualitative features of spatial patterns obtained for different lattice sizes remain consistent. Quantitatively, differences in stationary cooperation fractions fall within statistical errors. This demonstrates that lattice size does not alter the long-term steady-state conclusions within the parameter range considered in this work. Accordingly, only representative results for $L=100$ are presented here, while convergence tests for various lattice sizes are placed in the Response to Reviewers and are not plotted in the main manuscript.

\subsection{Effects of Initial Spatial Distributions on Steady-State Results}
To further test the robustness of our results against different initial spatial arrangements, we perform simulations with two distinct initial configurations. The first is the random initial condition, where cooperators and defectors are randomly and uniformly distributed over the entire lattice. The second is the block-clustered initial condition, in which cooperators are initially confined to the central region of the lattice while defectors occupy the peripheral area. We capture long-time spatial snapshots for four typical game classes (HG, SD, SH, PD), as shown in Fig. 8 and Fig. 9. Although these two initial conditions lead to remarkably different transient invasion and spreading dynamics in the early evolutionary stage, the asymptotic steady-state spatial structures converged by the system are identical for each game type. This result demonstrates that the initial spatial distribution of strategies only changes the transient evolutionary pathways and does not alter the final steady-state attractor of the spatial evolutionary system. In addition, consistent steady-state cooperation levels are obtained under different lattice sizes, which excludes spurious outcomes induced by finite-size effects and further verifies the reliability of the spatial dynamical features reported in this work.

\begin{figure}[htbp]
  \centering
  \includegraphics[width=1.0\textwidth]{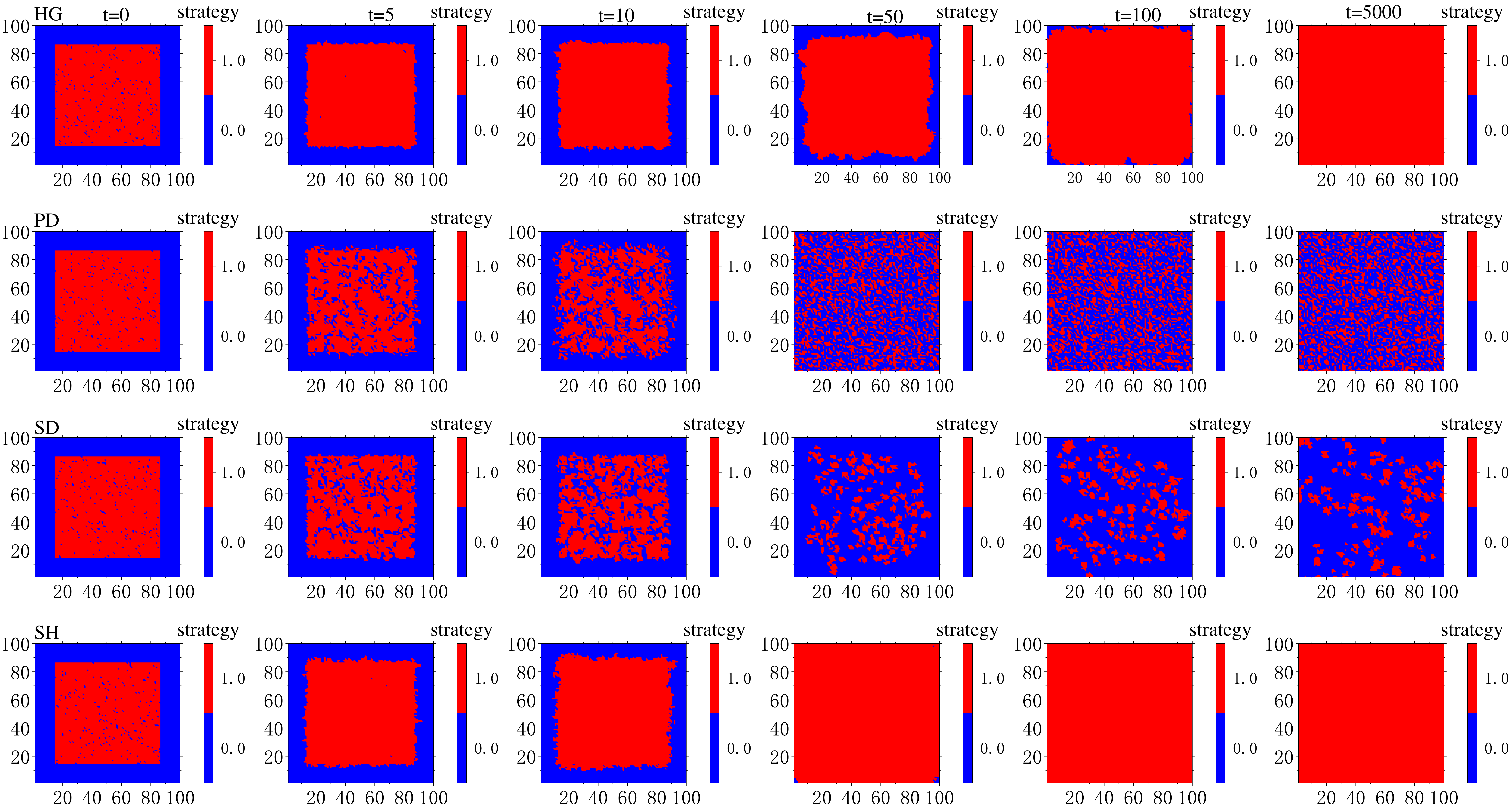}
  \caption{(Color online) Spatial evolutionary snapshots at representative Monte-Carlo time steps for four typical game types under the block-clustered initial configuration. Cooperators are colored red, defectors are colored blue.
  The four vertical rows correspond to different game types with fixed payoff parameters:
  (HG) Harmony Game: $R=3.0$, $S=1.0$, $P=0.0$, $T_{\text{val}}=2.0$;
  (PD) Prisoner's Dilemma: $R=3.0$, $S=1.0$, $P=2.0$, $T_{\text{val}}=4.0$;
  (SD) Snowdrift Game: $R=3.0$, $S=1.0$, $P=0.0$, $T_{\text{val}}=4.5$;
  (SH) Stag-Hunt Game: $R=3.0$, $S=1.0$, $P=2.0$, $T_{\text{val}}=1.0$.
  The time steps from left to right are: $t=0$, $t=5$, $t=10$, $t=50$, $t=100$, and $t=5000$.
  Other simulation parameters: lattice size $L=100$, periodic boundary conditions, Fermi-form logit imitation kernel with noise coefficient $w=\beta^{-1}=0.01$.}
  \label{fig.8}
\end{figure}

Figure \ref{fig.8} shows at the initial time \(t=0\) under the block-clustered initial configuration, cooperators (marked in red) are concentrated in the central square region of the lattice, while defectors (marked in blue) occupy the peripheral area. The horizontal subplots show spatial snapshots taken at Monte-Carlo time \(t=0,5,10,50,100,5000\), illustrating the evolutionary spatial patterns over time. The four vertical rows correspond to four game paradigms: the Harmony Game (HG), the Prisoner's Dilemma (PD), the Snowdrift Game (SD), and the Stag-Hunt Game (SH). For the Harmony Game, the initial central red cooperative block expands outward; strategy erosion occurs at the boundary in the early stage together with interface perturbations at intermediate time points, and the whole lattice turns fully red at \(t=5000\), indicating convergence to the full-cooperation steady state. For the Prisoner's Dilemma, the initial central red cooperative cluster gradually breaks and spreads; cooperators and defectors continuously interpenetrate each other without fully occupying the lattice, and a mixed spatial pattern with interleaved red and blue is maintained at \(t=5000\), corresponding to the statistical steady state of mixed strategies. For the Snowdrift Game, the central cooperative cluster spreads outward and splits into scattered cooperative fragments. After long-time evolution, numerous discrete red cooperative patches coexist with blue defector domains on the lattice, forming a mixed steady state with partial cooperation. For the Stag‑Hunt Game, the central cooperative cluster invades and expands persistently to compress defector domains. After a short transient period, the lattice is rapidly covered by red, and the system reaches the full-cooperation steady state at \(t=5000\).

\begin{figure}[htbp]
  \centering
  \includegraphics[width=1.0\textwidth]{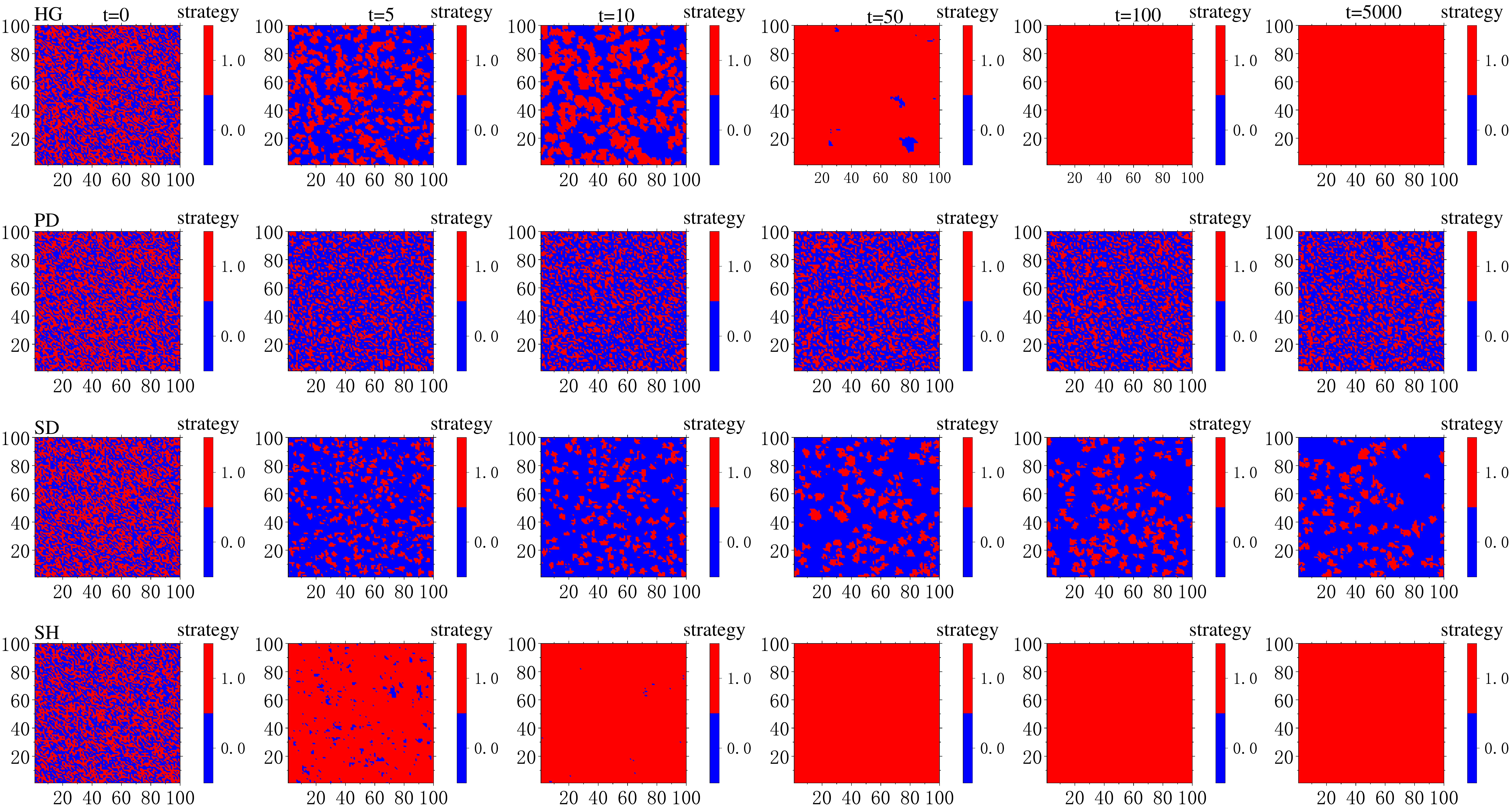}
  \caption{(Color online) Spatial evolutionary snapshots at representative Monte-Carlo time steps for four typical game types under the random initial configuration. Cooperators are colored red, defectors are colored blue.
  The four vertical rows correspond to different game types with fixed payoff parameters:
  (HG) Harmony Game: $R=3.0$, $S=1.0$, $P=0.0$, $T_{\text{val}}=2.0$;
  (PD) Prisoner's Dilemma: $R=3.0$, $S=1.0$, $P=2.0$, $T_{\text{val}}=4.0$;
  (SD) Snowdrift Game: $R=3.0$, $S=1.0$, $P=0.0$, $T_{\text{val}}=4.5$;
  (SH) Stag‑Hunt Game: $R=3.0$, $S=1.0$, $P=2.0$, $T_{\text{val}}=1.0$.
  The time steps from left to right are: $t=0$, $t=5$, $t=10$, $t=50$, $t=100$, and $t=5000$.
  Other simulation parameters: lattice size $L=100$, periodic boundary conditions, Fermi-form logit imitation kernel with noise coefficient $w=\beta^{-1}=0.01$.}
  \label{fig.9}
\end{figure}

Figure \ref{fig.9} shows at the initial time \(t=0\) under the random initial configuration, cooperators (red) and defectors (blue) are randomly and uniformly distributed over the whole lattice. The horizontal subplots present spatial snapshots at Monte-Carlo time \(t=0,5,10,50,100,5000\), depicting temporal evolutionary patterns. The four vertical rows correspond to the Harmony Game (HG), the Prisoner's Dilemma (PD), the Snowdrift Game (SD), and the Stag-Hunt Game (SH). For the Harmony Game, scattered cooperators gradually aggregate and expand, and the lattice becomes fully red at \(t=5000\), reaching the full-cooperation steady state. For the Prisoner's Dilemma, cooperators and defectors keep interpenetrating during evolution. Interleaved red-blue mixed patterns persist at \(t=5000\), representing the mixed-strategy statistical steady state. For the Snowdrift Game, isolated cooperators form scattered cooperative patches. These discrete red patches coexist with blue defector regions after long-time evolution and establish a partial-cooperation mixed steady state. For the Stag-Hunt Game, dispersed cooperators merge into large connected domains and suppress defector areas. The whole lattice turns red quickly within transient evolution, and the system arrives at the full-cooperation steady state at \(t=5000\).

\section{Conclusions}
\label{section.5}
This paper establishes a coupled strategy–payoff replicator dynamic framework to systematically unpack how payoff parameters modulate the evolutionary dynamics of cooperation. Combining rigorous mathematical derivation and multi-angle numerical verification, the core findings are summarized as follows.

First, we construct a two-strategy feedback-evolving game model and analytically solve all system fixed points together with their stability criteria. According to payoff ranking rules, the full parameter space is divided into four canonical game regimes (HG, SH, PD, SD). The two transcritical bifurcation thresholds $T_{c1}(P)$ and $T_{c2}(P)$ are functions of the punishment parameter $P$. At $T_{c1}(P)$, the full-cooperation equilibrium loses stability, while $T_{c2}(P)$ triggers stability inversion of interior mixed equilibria. As defection temptation $T$ rises, the system continuously shifts between distinct game regimes, proving that $T$ acts as the dominant factor governing game categories and equilibrium configurations. Each regime bears unique evolutionary features: HG displays strong bistability sensitive to initial cooperation fractions; SH sustains bistability where high initial prevalence safeguards universal cooperation; SD exhibits bistable dynamics with coexisting boundary and interior equilibria, while PD converges to a unique persistent mixed coexistence equilibrium, embodying the core conflict of social dilemmas. All theoretical predictions on fixed-point stability, threshold bifurcation and regime transitions are fully validated via $\mathrm{d}x/\mathrm{d}t$ phase portraits and time-series trajectory simulations.

Second, two-dimensional lattice simulations with periodic boundaries are implemented to capture spatial interaction effects. Spatial clustering eliminates the absorbing full-defection state and maintains non-zero cooperation across all game types. Within HG and SH, lattice topology stabilizes full cooperation and speeds up convergence to cooperative equilibria. For SD and PD, spatial structures produce sustained fluctuating mixed states whose average cooperation share rises with initial cooperator density. The qualitative consistency between lattice simulation outputs and mean-field replicator results verifies the broad applicability and robustness of our theoretical framework. The unified bifurcation analysis paradigm developed herein offers standardized theoretical and methodological references for studying cooperation in high-dimensional replicator systems and networked evolutionary games.

This work contains several idealized simplifications that restrict its direct application to real complex systems: the baseline mean-field model assumes infinite well-mixed populations and overlooks individual heterogeneity and localized interactions; only binary pure strategies (cooperation/defection) are considered, excluding multi-strategy competition and endogenous update rule evolution; deterministic replicator dynamics are adopted without stochastic noise or adaptive learning behaviors. Although supplementary lattice simulations introduce spatial interactions, they still rely on regular grid topology and pairwise two-player games. Future research will extend this framework to heterogeneous complex networks, stochastic imitation rules, individual learning dynamics and multi-layer social interaction architectures for higher real-world fidelity.

\section*{Declaration of Competing Interest}
Authors declare that they have no known competing financial interests or personal relationships that could have appeared to influence the work reported in this paper.

\section*{CRediT authorship contribution statement}
\textbf{Yu Chen}: Writing - original draft, editing, Visualization.
\textbf{Ji-Qiang Zhang}: Writing - review \& editing, Validation.
\textbf{Bin-Quan Li}: Writing - review \& editing, Validation, Supervision, Project administration, Funding acquisition.

\section*{Data availability}
Data will be made available on request.

\section*{Acknowledgment}
We are supported by the National Natural Science Foundation of China under Grant No. 12405040, the Key
Research and Development Program of Ningxia Province in China under Grant No. 2025BEH04042 and the Natural Science Foundation of Ningxia Province in China under Grant No. 2026AAC030077.

\bibliography{citation}

\end{document}